\documentclass[review]{elsarticle}
\makeatletter
\def\ps@pprintTitle{%
 \let\@oddhead\@empty
 \let\@evenhead\@empty
 \def\@oddfoot{\centerline{\thepage}}%
 \let\@evenfoot\@oddfoot}
\makeatother

\usepackage{lineno,hyperref}
\usepackage{graphicx}      
\usepackage{natbib}        
\usepackage{amsmath}
\usepackage{amssymb}
\usepackage{float}
\usepackage{algorithm}
\usepackage{algpseudocode}

\usepackage{setspace}

\begin{document}

\begin{frontmatter}

\title{Learning Linear Representations of Nonlinear Dynamics Using Deep Learning}

\author[1]{Akhil Ahmed}
\author[1]{Ehecatl Antonio del Rio-Chanona}
\author[1]{Mehmet Mercangöz\corref{cor1}}
\ead{m.mercangoz@imperial.ac.uk}

\cortext[cor1]{Corresponding author}
\address[1]{Department of Chemical Engineering, Imperial College London, London, UK}

\begin{abstract}
The vast majority of systems of practical interest are characterised by nonlinear dynamics. This renders the control and optimization of such systems a complex task due to their nonlinear behaviour. Additionally, standard methods such as linearizing around a fixed point may not be an effective strategy for many systems, thus requiring an alternative approach. For this reason, we propose a new deep learning framework to discover a transformation of a nonlinear dynamical system to an equivalent higher dimensional linear representation. We demonstrate that the resulting learned linear representation accurately captures the dynamics of the original system for a wider range of conditions than standard linearization. As a result of this, we show that the learned linear model can subsequently be used for the successful control of the original system. We demonstrate this by applying the proposed framework to two examples; one from the literature and a more complex example in the form of a Continuous Stirred Tank Reactor (CSTR). 
\end{abstract}


\end{frontmatter}

\section{Introduction}\label{introduction}
Nonlinear systems are prevalent in nature with most systems of practical interest exhibiting nonlinear behaviour. For this reason, the control and optimization of such systems is a vitally important task. However, at the same time, this task is a formidable challenge with no general and scalable solution. This is especially relevant for large-scale systems, the like of which are often encountered in systems engineering, which encompass multiscale spatiotemporal phenomena which can be difficult to accurately model. Moreover, even if such a model could be identified, the resulting model structure may be too complex for the tractable control and optimization of the system of interest \citep{Proctor2018}, \cite{Siljak2005}, \cite{Antoulas2005}. 

In contrast, the study of linear systems is well developed with scalable design, analysis, control and optimization of linear systems thoroughly detailed within the literature \cite{Overschee1996}, \cite{Ogata2009}. To take advantage of these developments, one can obtain linear approximations of nonlinear systems by linearizing around an operating point. While this may prove to be an effective strategy for some nonlinear systems, this may not be generally applicable especially for systems exhibiting strong nonlinearities. Instead, for such systems, a more general approach is required. In such instances, it may be possible to consider a coordinate transformation or a change of variables such that the original nonlinear system is transformed into an equivalent linear system. For example, such ideas are leveraged in feedback linearization to control nonlinear systems \cite{Brockett1983}, \cite{Rubio2018}.

Along these lines, there has been renewed interest in the seminal work of Bernard Koopman on Koopman operator theory \cite{Koopman1931}, \cite{Banaszuk2004}, \cite{Mezic2005}. In essence, Koopman operator theory allows for a nonlinear system to be represented as an infinite dimensional linear system. This is done by considering transformations of the original system variables such that the resulting system is linear in the new variables. This is discussed in more detail and demonstrated with an illustrative example in section \ref{background}.

It could be argued that Koopman operator theory does not solve the original problem as nonlinearity is traded for infinite dimensionality. However, with the advent of more powerful data-driven methods in machine learning and system identification, finite dimensional approximations can be obtained of the infinite dimensional linear representation. In particular, Dynamic Mode Decomposition (DMD) was proposed and further developed by the authors of \cite{Rowley2009} and \cite{Schmid2010} to compute these finite dimensional approximations. Consequently, DMD, and by extension, Koopman operator theory, has been applied with great success to a broad range of fields, ranging from fluid mechanics to neuroscience, all of which are underpinned by nonlinear dynamics \cite{Brunton2016}, \cite{Schmid2011}.

Despite this success, DMD is restricted to linear transformations of the original system variables which renders the approach restrictive for many nonlinear systems of interest \cite{Kaiser2021}. Consequently, to allow for a richer set of transformations many alternative approaches have been proposed such as the use of kernel functions to neural networks \cite{Williams2014}, \cite{Lusch2018}. Additionally, Koopman operator theory, in its original form, does not consider the effects of inputs and control for the nonlinear system. However, in recent years, many advances have been made to generalise Koopman operator theory to allow for the action of inputs and the resulting control of the system under study \cite{Proctor2018}, \cite{Kaiser2021}, \cite{Lian2021}. Consequently, these extensions could be used to represent complex nonlinear systems with finite dimensional linear approximations which could subsequently be used for the tractable control of the system. 

For the above reasons, in this paper, we propose a deep learning framework to discover a transformation of a nonlinear system to an equivalent higher dimensional linear system. In the proposed framework, the neural network serves the role of parameterizing the function space over which the transformation is searched for, allowing for a tractable optimization problem to be solved. Additionally, we leverage both forward and backpropagation to relate the original dynamics to the learned transformed dynamics by using the chain rule to approximate the Jacobian of the transformation. Moreover, we demonstrate that the use of neural networks, through the collection of activation functions, allows for a broader range of transformations of the original system variables to be considered. Consequently, through simulation results we are able to accurately capture the original system dynamics using the learned linear representation for a wider range of operating conditions than would be possible with standard linearization. Additionally, we use the resulting linear representations to successfully control the original system. 

This paper is outlined as follows: In section \ref{background}, a brief background of Koopman operator theory is provided, aided with a simple illustrative example. This is followed by an introduction to the proposed deep learning framework in section \ref{problem_formulation}. Finally, the proposed method is applied to two examples in section \ref{results}: a commonly cited example from the literature and a more complex example in the form of a Continuous Stirred Tank Reactor (CSTR). Challenges and future directions of the proposed approach are also discussed in section \ref{results}. Finally, the paper is concluded in section \ref{conclusions}.

\section{Background}\label{background}

In this section, a brief background on Koopman operator theory is provided, motivated by an illustrative example. For a more thorough treatment, the reader is directed to the works of \cite{Proctor2018}, \cite{Mezic2005} and \cite{Kaiser2021}.

\subsection{Koopman Operator Theory} \label{KOT}

Consider a nonlinear dynamical system defined by:

\begin{equation}\label{nonlinear_dynamics}
\frac{\mathrm{d}\mathbf{x}}{\mathrm{d}t} = \mathbf{f(x)}
\end{equation}
where $\mathbf{x} \in \mathbb{R}^{n_x}$ and $ \mathbf{f}: \mathbb{R}^{n_x} \to \mathbb{R}^{n_x}$. To demonstrate the action of the Koopman operator, we can equivalently consider the dynamics in discrete time as:

\begin{equation}\label{nonlinear_dynamics_discrete}
\mathbf{x}_{k+1} = \mathbf{F}(\mathbf{x}_k)
\end{equation}
where $\mathbf{x}_{k+1},\mathbf{x}_k  \in \mathbb{R}^{n_x}$ while $ \mathbf{F}: \mathbb{R}^{n_x} \to \mathbb{R}^{n_x}$ defines an update to advance the state forward one step from $\mathbf{x}_k$ to $\mathbf{x}_{k+1}$. 

In his seminal paper, Koopman defined a set of ``measurement" functions or transformations of the original state variables defined by $ g:\mathbb{R}^{n_x} \to \mathbb{R} $. He then demonstrated that there exists an infinite dimensional linear operator, which we refer to as the Koopman operator, $\mathcal{K}$, which acts on this set of measurement functions to advance the measurements linearly in time:

\begin{equation}\label{koopman_def}
\begin{aligned}
g(\mathbf{F}(\mathbf{x}_k)) &= \mathcal{K}g(\mathbf{x}_k)\\
g(\mathbf{x}_{k+1}) &= \mathcal{K}g(\mathbf{x}_k)
\end{aligned}
\end{equation}

As discussed in section \ref{introduction}, although the Koopman operator is linear, it is also infinite dimensional thus requiring an approach (such as DMD) to generate a finite dimensional approximation. As discussed in section \ref{problem_formulation}, in this paper we will demonstrate how this can be achieved using deep learning. 

To motivate the idea behind Koopman operator theory, we will illustrate the concept with a simple example in section \ref{motivating_example}.

\subsection{A Motivating Example} \label{motivating_example}

Consider the nonlinear dynamical system defined by:

\begin{equation}\label{simple_example}
\frac{\mathrm{d}x}{\mathrm{d}t} = x^{2}
\end{equation}
where $x \in \mathbb{R}$. As per section \ref{KOT}, we wish to find a transformation of state $x$ such that in the new system representation, the dynamics are characterised by a linear system. In this case, a simple, closed form solution exists defined by the transformation:

\begin{equation}\label{simple_transformation}
z = e^{-\frac{1}{x}}
\end{equation}

Using the chain rule, we can verify that equation (\ref{simple_transformation}) defines a transformation which converts the original system into a linear system while satisfying the original dynamics:

\begin{equation}\label{simple_linear}
\begin{split}
\frac{\mathrm{d}z}{\mathrm{d}t} & = \frac{\mathrm{d}z}{\mathrm{d}x}\frac{\mathrm{d}x}{\mathrm{d}t}\\
             & = e^{-\frac{1}{x}}\\
             & = z
\end{split}
\end{equation}

Consequently, as per equation (\ref{simple_linear}), the original dynamics are now linear in the new state, $z$, defined by the transformation given in equation (\ref{simple_transformation}). 

\paragraph{Remarks} It should be noted that this example is only intended to motivate the idea behind Koopman operator theory. It is a trivial example for which a closed form solution can be found easily. Additionally, notice that the transformation from state $x$ to $z$ stays within the same dimension. This will not be generally true for most systems of interest which are characterised by complex nonlinear dynamics. In such cases, it will likely be necessary to ``lift" the dynamics to a higher dimension. Consequently, a more general approach is required to discover these transformations. Additionally, we have not yet considered the impact of inputs on the system dynamics. It is for these reasons that we propose a deep learning framework to achieve these goals as discussed in the next section, section \ref{problem_formulation}.

\section{Problem Formulation}\label{problem_formulation}
In this section, we will introduce the proposed deep learning framework. In section \ref{nonlinear_systems}, we will start by formulating the problem applied to the case of nonlinear dynamics with no inputs. In section \ref{nonlinear_systems_w_inputs}, we will extend this to systems with exogenous inputs. Finally, in section \ref{solve_ocp} we will demonstrate how the learned linear representation from section \ref{nonlinear_systems_w_inputs} can be used to solve the optimal control problem to control the original nonlinear system.

\subsection{Nonlinear Systems}\label{nonlinear_systems}
Once again, consider a nonlinear dynamical system defined by:

\begin{equation}\label{nonlinear_dynamics}
\frac{\mathrm{d}\mathbf{x}}{\mathrm{d}t} = \mathbf{f(x)}
\end{equation}
where $\mathbf{x} \in \mathbb{R}^{n_x}$ and $ \mathbf{f}: \mathbb{R}^{n_x} \to \mathbb{R}^{n_x}$. As per the discussions of section \ref{background}, we wish to seek a transformation defined by:

\begin{equation}\label{transformation}
\boldsymbol{\phi} : \mathbb{R}^{n_x} \to \mathbb{R}^{n_z}
\end{equation}
where $n_x$ is the dimension of the original system and $n_z$ is the dimension of the new transformed system where $n_z \geq n_x$. Additionally, we define the new state variable after the transformation as:

\begin{equation}\label{new_var}
\mathbf{z} = \boldsymbol{\phi(\mathbf{x})}
\end{equation}
where $\mathbf{z} \in \mathbb{R}^{n_z}$. From the chain rule, the relationship between the new state, $\mathbf{z}$, and the original state, $\mathbf{x}$, can be defined:

\begin{equation}\label{chain}
\frac{\mathrm{d}\mathbf{z}}{\mathrm{d}t} = \mathbf{J_z(x)}\frac{\mathrm{d}\mathbf{x}}{\mathrm{d}t} 
\end{equation}
where $\mathbf{J_z(x)}\in \mathbb{R}^{n_z \times n_x}$ is the Jacobian matrix of the transformation, $\boldsymbol{\phi}$. Crucially, we seek to satisfy the condition that the dynamics of the new state $\mathbf{z}$ are linear in $\mathbf{z}$:

\begin{equation}\label{linear}
\frac{\mathrm{d}\mathbf{z}}{\mathrm{d}t} = \mathbf{A}\mathbf{z}
\end{equation}
where $\mathbf{A} \in \mathbb{R}^{n_z \times n_z}$. Combining equations (\ref{chain}) and (\ref{linear}) defines a system of partial differential equations for the the transformation $\mathbf{z} = \boldsymbol{\phi(\mathbf{x})}$:

\begin{equation}\label{PDE_sys}
\frac{\mathrm{d}\mathbf{z}}{\mathrm{d}t} = \mathbf{J_z(x)}\frac{\mathrm{d}\mathbf{x}}{\mathrm{d}t}  = \mathbf{A}\mathbf{z}
\end{equation}

Notice that, in general, equation (\ref{PDE_sys}) defines an underdetermined system of $n_z$ equations with $n_z$ unknown transformations and $n_z \cdot n_z$ unknown coefficients in matrix $\mathbf{A}$. However, for some special cases, a closed form solution can be found directly. An example of this is the motivating example from section \ref{motivating_example}, where $n_x = n_z = 1$, and as such an ordinary differential equation is recovered which can be solved using standard methods. In general, however, equation (\ref{PDE_sys}) can not be solved directly. 

Alternatively, the problem can be solved by reformulation of equation (\ref{PDE_sys}) into an optimization problem which can be solved using data measured from the system. Specifically, an objective function can be defined as the squared Euclidean norm of the difference between equation (\ref{chain}) and (\ref{linear}):

\begin{equation}\label{objective}
J = \left\|\mathbf{J_z(x)}\frac{\mathrm{d}\mathbf{x}}{\mathrm{d}t}  -  \mathbf{A}\mathbf{z}\right\|_2^2
\end{equation}
where $J \in \mathbb{R}$ and $||\cdot||_2: \mathbb{R}^{n_z} \to \mathbb{R}$. Recalling equation (\ref{new_var}), the unconstrained optimization problem can then be defined as an optimization over functions, $\boldsymbol{\phi}$, and parameters in matrix $\mathbf{A}$:

\begin{equation}\label{optimisation_one}
\underset{\boldsymbol{\phi},\mathbf{A}}{\text{min}} \; \left\|\mathbf{J_\mathbf{\boldsymbol{\phi}}(x)}\frac{\mathrm{d}\mathbf{x}}{\mathrm{d}t}  -  \mathbf{A}\boldsymbol{\phi(\mathbf{x})}\right\|_2^2
\end{equation}

However, for most practical problems of interest, optimization over function spaces is generally intractable. Therefore, an alternative approach would be to parameterize the space of functions or to have a set of basis functions from which the space of functions can be generated \cite{Sasane2016}. This is the approach adopted in the current work. In particular, neural networks can be regarded as a parametrization of functions where the collection of activation functions define the set of basis functions while the weights of the neural network serve as the parameters. As a result, we use the neural network to parameterize the transformation defined in equation (\ref{transformation}) and define this as below:

\begin{equation}\label{param_func}
\mathbf{z} = \boldsymbol{\phi(\mathbf{x})} = \mathbf{N(x;p)}
\end{equation}
where $\mathbf{p} \in \mathbb{R}^{n_p}$ is the vector of neural network parameters (weights) and $n_p$ is the total number of weights in the network. $\mathbf{N}: \mathbb{R}^{n_x} \times \mathbb{R}^{n_p} \to \mathbb{R}^{n_z}$ defines the neural network which transforms the original state, $\mathbf{x}$, to the new state, $\mathbf{z}$, given parameters, $\mathbf{p}$. As a result, the optimization problem defined in equation (\ref{optimisation_one}) can be reformulated into a tractable form as an optimization purely over parameters alone:

\begin{equation}\label{optimisation_two}
\underset{\boldsymbol{p},\mathbf{A}}{\text{min}} \; \left\|\mathbf{J_\mathbf{\boldsymbol{N}}(x)}\frac{\mathrm{d}\mathbf{x}}{\mathrm{d}t} -  \mathbf{A}\mathbf{N(x;p)}\right\|_2^2
\end{equation}

At this point, it should be noted that the idea of using neural networks to solve differential equations (such as equation \ref{PDE_sys}) is not new. It was first proposed by the authors of \cite{Lagaris1998} and then popularized by Physics Informed Neural Networks (PINNs) as per the work of \cite{Raissi2019}. Despite this, it is not a typical task which is usually ascribed to neural networks. As a result of this, stability issues were encountered during training, some of which are documented in the literature \cite{Wang2021}. As an example, notice that equation (\ref{optimisation_two}) is trivially minimized by the solution, $\mathbf{p} = \mathbf{0}$ and $\mathbf{A} = \mathbf{0}_{n_z \times n_z}$. To avoid such issues, an additional term was added to the objective function to re-frame the problem as a regression task for the neural network. In essence, the additional term defines a ``decoder" network to perform the inverse transformation from the new state, $\mathbf{z}$, back to a reconstruction of $\mathbf{x}$, which we denote as $\mathbf{\hat{x}}$ \cite{Hinton2006}. This results in the additional objective term defined in equation (\ref{decoder}):

\begin{equation}\label{decoder}
J_{\mathrm{decoder}} = \left\|\mathbf{x} - \mathbf{\hat{x}}\right\|_2^2
\end{equation}

The overall objective function then consists of a weighted sum of the two terms giving the optimization problem defined in equation (\ref{full_opt}):

\begin{equation}\label{full_opt}
\underset{\boldsymbol{p},\mathbf{A},\mathbf{q}}{\text{min}} \; \left\|\mathbf{J_\mathbf{\boldsymbol{N}}(x)}\frac{\mathrm{d}\mathbf{x}}{\mathrm{d}t} -  \mathbf{A}\mathbf{N(x;p)}\right\|_2^2
+
\lambda\left\|\mathbf{x} - \mathbf{\hat{x}}\right\|_2^2
\end{equation}
where $\mathbf{q} \in \mathbb{R}^{n_q}$ is the vector of decoder network parameters while $\lambda$ defines a scalar weighting factor applied to the decoder term. Consequently, the optimization problem is solved as per Algorithm \ref{algo1}.

\begin{algorithm}[H]
\setstretch{1.1}
\caption{Algorithm for solving equation (\ref{full_opt})}\label{algo1}
\begin{algorithmic}[1]
\Require initial weighting factor $\lambda_{initial}$; decay rate $\gamma > 1$
\Ensure optimal parameters $\Theta = (\mathbf{p},\mathbf{A},\mathbf{q})$
\State $\lambda \gets \lambda_{initial}$
\State Initialize parameters as vectors/matrix sampled from standard normal distribution: $\mathbf{p} \sim \mathcal{N}(\mathbf{0},\mathbf{1}) \in \mathbb{R}^{n_p}$; $\mathbf{A} \sim \mathcal{N}(\mathbf{0},\mathbf{1}) \in \mathbb{R}^{n_z \times n_z}$; $\mathbf{q} \sim \mathcal{N}(\mathbf{0},\mathbf{1}) \in \mathbb{R}^{n_q}$
\State $\Theta \gets (\mathbf{p},\mathbf{A},\mathbf{q})$
\While{termination criteria not met}
\State $\hat{\Theta} \gets \text{arg}\underset{\boldsymbol{p},\mathbf{A},\mathbf{q}}{\text{min}} \; \left\|\mathbf{J_\mathbf{\boldsymbol{N}}(x)}\frac{\mathrm{d}\mathbf{x}}{\mathrm{d}t} -  \mathbf{A}\mathbf{N(x;p)}\right\|_2^2
+
\lambda\left\|\mathbf{x} - \mathbf{\hat{x}}\right\|_2^2$
\State $\Theta \gets \hat{\Theta}$
\State $\lambda \gets \lambda/\gamma$
\EndWhile
\State
\Return $\Theta$
\end{algorithmic}
\end{algorithm}

A brief summary of the steps is presented here:

\textbf{Step 1}: The weighting factor, $\lambda$, applied to the decoder term in  equation (\ref{full_opt}), is initialized with a large weight, $\lambda_{initial}$.

\textbf{Step 2}: The decision variables, $(\mathbf{p,A,q})$, are initialized as random vectors/matrix sampled from the appropriate multivariate standard normal distribution. 

\textbf{Step 3}: The parameters from step 2 are initialized as the current set of ``optimal" parameters.

\textbf{Step 4}: A while loop is initiated until one of the termination criteria are satisfied. That is, either the optimal parameters between iterations are similar, where similarity is measured with a Euclidean distance metric, or the maximum number of iterations are exceeded. 

\textbf{Step 5}: Equation (\ref{full_opt}) is solved either to local optimality or until the maximum number of optimization iterations are exceeded. The optimal parameters are assigned to the variable, $\hat{\Theta}$. The optimization problem for the examples considered in this paper were solved using IPOPT \cite{Wachter2006}. 

\textbf{Step 6}: Provided the first termination criteria is not satisfied i.e. the optimal parameters between iterations are not similar, then the current set of optimal parameters are updated.

\textbf{Step 7}: The weighting factor, $\lambda$, is divided by the decay rate parameter, $\gamma$.

\textbf{Step 8}: The while loop is terminated when one of the termination criteria are satisfied. 

\textbf{Step 9}: The optimal set of parameters, $\Theta = (\mathbf{p,A,q})$, are returned.

\begin{figure}
\begin{center}
\includegraphics[width=0.8\textwidth]{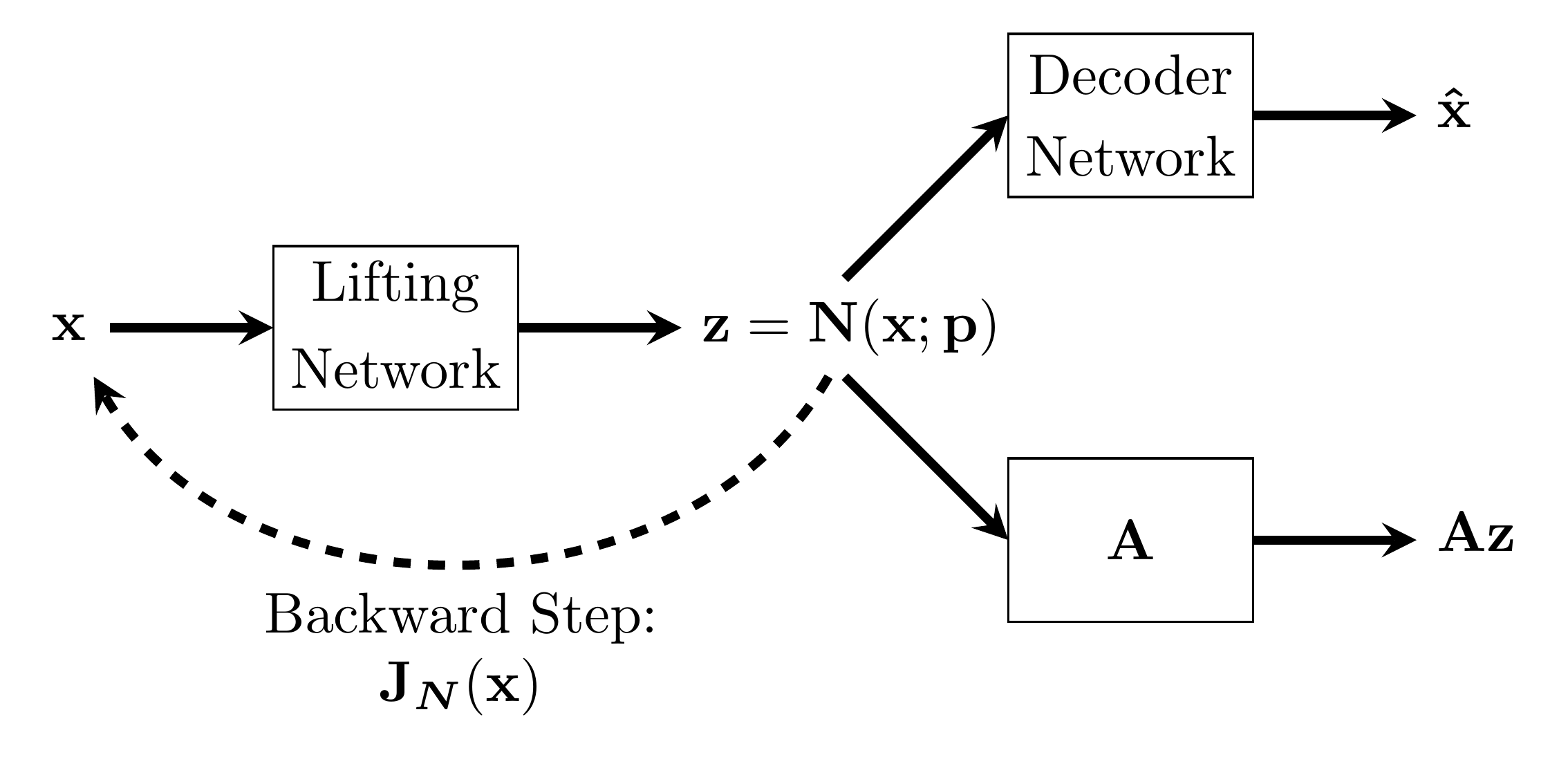} 
\caption{
Schematic diagram of the proposed deep learning framework. The original state, $\mathbf{x}$, is transformed via the lifting neural network to obtain the new state, $\mathbf{z}$, which is then subjected to a linear transformation by the matrix $\mathbf{A}$. The Jacobian matrix of the transformation, $\mathbf{J_\mathbf{\boldsymbol{N}}(x)}$, can be obtained via a backward step through the lifting network. Finally, a decoder network is used to transform the new state, $\mathbf{z}$, back to a reconstruction of the original state, $\mathbf{\hat{x}}$.
} 
\label{network_set_up}
\end{center}
\end{figure}

While the details aforementioned are important to consider, in general, the objective of the proposed framework is to solve the optimization problem formulated in equation (\ref{optimisation_two}). Consequently, the rest of the paper will proceed with this implicit understanding in mind. For reference, Fig. \ref{network_set_up} provides a visualization of the entire set-up.

With the optimization problem formulated as equation (\ref{optimisation_two}), we can now solve the problem using data measured from the system. This can be done by collecting measurements of the original state, $\mathbf{x}$, and its time derivative, $\frac{\mathrm{d}\mathbf{x}}{\mathrm{d}t}$, through time. Consequently, each of the terms in the objective function defined in equation (\ref{optimisation_two}) can be calculated: $\mathbf{N(x;p)}$ can be determined by a forward pass through the network while the Jacobian term, $\mathbf{J_\mathbf{\boldsymbol{N}}(x)}$, can be determined via a backward step through the network. This is done for each sample in the dataset giving the final optimization problem defined in equation (\ref{optimisation_sum}), where $N$ is the number of samples collected and the superscript defines the $i$th sample.

\begin{equation}\label{optimisation_sum}
\underset{\boldsymbol{p},\mathbf{A}}{\text{min}} \; \sum_{i=0}^{N}\left\|\mathbf{J}_\mathbf{\boldsymbol{N}}(\mathbf{x}^{(i)})\left(\frac{\mathrm{d}\mathbf{x}}{\mathrm{d}t}\right)^{(i)} -  \mathbf{A}\mathbf{N}(\mathbf{x}^{(i)};\mathbf{p})\right\|_2^2
\end{equation}

In cases where obtaining direct measurements of the time derivative are infeasible, it can alternatively be approximated from measurements of $\mathbf{x}$. For noisy measurements, total variation regularization can be used to approximate the derivative as discussed in \cite{Steve2016}.

For the examples considered in section \ref{results} of this paper, all data was collected directly from simulation results. 

\subsection{Nonlinear Systems with Inputs}\label{nonlinear_systems_w_inputs}

Up until now, we have only considered nonlinear systems without any inputs acting on the system. In this section, we generalise this to nonlinear systems with exogenous inputs with a focus on control-affine systems for reasons which will be discussed shortly. 

Consider a general nonlinear dynamical system with inputs defined by:

\begin{equation}\label{nonlinear_dynamics_input}
\frac{\mathrm{d}\mathbf{x}}{\mathrm{d}t}  = \mathbf{f(x,u)}
\end{equation}
where $\mathbf{x} \in \mathbb{R}^{n_x}$, $\mathbf{u} \in \mathbb{R}^{n_u}$ and $ \mathbf{f}: \mathbb{R}^{n_x} \times \mathbb{R}^{n_u} \to \mathbb{R}^{n_x}$. As before, we seek a transformation defined by:

\begin{equation}\label{transformation_inputs}
\boldsymbol{\phi} : \mathbb{R}^{n_x} \times \mathbb{R}^{n_u} \to \mathbb{R}^{n_z}
\end{equation}
where $n_x$ is the number of states of the original system, $n_u$ is the number of inputs to the original system while $n_z$ is the number of states of the new transformed system with $n_z \geq (n_x + n_u)$. Notice that the transformations defined by equation (\ref{transformation_inputs}) may include state-input cross-terms (for example, terms such as $x_iu_j$). This complicates control for the new system, as the new state being controlled may depend on an input from the original system. This is discussed by the authors of \cite{Proctor2018} and \cite{Kaiser2021}.

For this reason, we focus our attention on a specific class of nonlinear systems where the effects of the states and inputs are decoupled: control-affine systems. Note that the examples considered in section \ref{results}, are restricted to control-affine systems for this reason. 

A control-affine system is defined as below:

\begin{equation}\label{control_affine}
\frac{\mathrm{d}\mathbf{x}}{\mathrm{d}t}  = \mathbf{f(x)} + \mathbf{Bu}
\end{equation}
where $\mathbf{x} \in \mathbb{R}^{n_x}$, $\mathbf{u} \in \mathbb{R}^{n_u}$, $\mathbf{B} \in \mathbb{R}^{n_x \times n_u}$ and $ \mathbf{f}: \mathbb{R}^{n_x} \to \mathbb{R}^{n_x}$. As dicussed in \cite{Kaiser2021}, due to the decoupling of the states and inputs, we can now restrict ourselves to finding a transformation of the states alone:

\begin{equation}\label{transformation_caf}
\boldsymbol{\phi} : \mathbb{R}^{n_x} \to \mathbb{R}^{n_z}
\end{equation}
where, as before, $n_x$ is the dimension of the original system and $n_z$ is the dimension of the new transformed system where $n_z \geq n_x$. We define the new state variable as before:

\begin{equation}\label{new_var_caf}
\mathbf{z} = \boldsymbol{\phi(\mathbf{x})}
\end{equation}

Applying the chain rule, we recover:

\begin{equation}\label{chain_caf}
\frac{\mathrm{d}\mathbf{z}}{\mathrm{d}t}  = \mathbf{J_z(x)}\frac{\mathrm{d}\mathbf{x}}{\mathrm{d}t}
\end{equation}
where all variables are defined as they were previously. As in section \ref{nonlinear_systems}, we wish to satisfy the condition that the new state dynamics are linear:

\begin{equation}\label{linear_caf}
\frac{\mathrm{d}\mathbf{z}}{\mathrm{d}t}  = \mathbf{A}\mathbf{z}
\end{equation}

By expanding equation (\ref{chain_caf}) through substitution of equation (\ref{control_affine}) for the time derivative term, and accounting for the condition we wish to satisfy via equation (\ref{linear_caf}) we obtain:

\begin{equation}\label{expanded_caf}
\frac{\mathrm{d}\mathbf{z}}{\mathrm{d}t}  = \mathbf{A}\mathbf{z} + \mathbf{J_z(x)}\mathbf{Bu}
\end{equation}

Notice that equation (\ref{expanded_caf}) is linear in the states, $\mathbf{z}$, but not necessarily jointly linear in the inputs and states due to the Jacobian term, $\mathbf{J_z(x)}$, which may be state dependent. Although the resulting system is not fully linear, as discussed in \cite{Kaiser2021}, the state-dependent control term, $\mathbf{J_z(x)B}$, does not pose any major issues with regards to control. This is discussed in more detail in section \ref{solve_ocp}, where we explain how equation (\ref{expanded_caf}) can be used to solve the optimal control problem to control the original nonlinear dynamical system, equation (\ref{control_affine}).

As per section \ref{nonlinear_systems}, we can combine equations (\ref{chain_caf}) and (\ref{expanded_caf}) and once again solve the problem of discovering transformations of the form of equation (\ref{transformation_caf}) as an optimization problem. As before, we parameterize the transformation using a neural network, as per equation (\ref{param_func}), which allows us to define the following optimization problem:

\begin{equation}\label{optimisation_control}
\underset{\boldsymbol{p},\mathbf{A}}{\text{min}} \; \left\|\mathbf{J_\mathbf{\boldsymbol{N}}(x)}\frac{\mathrm{d}\mathbf{x}}{\mathrm{d}t}  -  \mathbf{A}\mathbf{N(x;p)-\mathbf{J_N(x)}\mathbf{Bu}}\right\|_2^2
\end{equation}

As explained in section \ref{nonlinear_systems}, the optimization problem can now be solved provided that data measured from the system is available. In the case of equation (\ref{optimisation_control}), additional data in the form of the input measurements, $\mathbf{u}$, through time is necessary. This gives the final optimization problem defined over the entire dataset:

\begin{equation}\label{optimisation_control_sum}
\underset{\boldsymbol{p},\mathbf{A}}{\text{min}} \; \sum_{i=0}^{N}\left\|\mathbf{J}_\mathbf{\boldsymbol{N}}(\mathbf{x}^{(i)})\left(\frac{\mathrm{d}\mathbf{x}}{\mathrm{d}t}\right)^{(i)} - \mathbf{A}\mathbf{N}(\mathbf{x}^{(i)};\mathbf{p})-\mathbf{J}_\mathbf{\boldsymbol{N}}(\mathbf{x}^{(i)})\mathbf{B}\mathbf{u}^{(i)}\right\|_2^2
\end{equation}
where $N$ is the number of samples collected and the superscript defines the $i$th sample. Additionally, for the same reasons as discussed in section \ref{nonlinear_systems}, Algorithm \ref{algo1} is again used to solve equation (\ref{optimisation_control_sum}) only with the new optimization problem replacing step 5 of the algorithm.

\subsection{Solving the Optimal Control Problem}\label{solve_ocp}

In this section, we discuss how the optimal control problem can be solved to control nonlinear dynamical systems of the form of equation (\ref{control_affine}), using the learned linear representation of the form of equation (\ref{expanded_caf}). This is primarily based on the work of \cite{Kaiser2021}.

Firstly, to motivate this section, allow us to reexamine the purpose of discovering a linear representation of the original nonlinear dynamics. As discussed in section \ref{introduction}, the study of linear systems is thoroughly developed. In particular, powerful and efficient methods exist for the optimal control of linear systems. In contrast, the study of nonlinear systems is less developed and no such general and scalable methods exist. Therefore, by learning a linear representation of the original nonlinear dynamics, the aforementioned methods for linear systems can be applied with ease \cite{Korda2018}.

However, as noted in section \ref{nonlinear_systems_w_inputs}, the learned transformation from the proposed framework i.e. equation (\ref{expanded_caf}) is not necessarily jointly linear in the inputs and states due to the potentially state-dependent control term, $\mathbf{J_z(x)B}$. As this term may be state-dependent, then linear optimal control can not be applied directly as the potential nonlinearity posed by this term must be managed. In contrast, if the learned transformation resulted in a fully linear system then a closed form solution exists to the optimal control problem. This is found by solving the Riccati equation for the optimal gain matrix, $\mathbf{K} \in \mathbb{R}^{n_u \times n_z}$, giving the optimal control law:

\begin{equation}\label{optimal_controller}
\mathbf{u} = - \mathbf{K}\mathbf{z}
\end{equation}

This is referred to as the Linear Quadratic Regulator (LQR) \cite{Ogata2009}. Consequently, the optimal control law, of the form of equation (\ref{optimal_controller}), could be directly applied to the original system.

However, as discussed, in the current work, the discovered transformation results in a model of the form of equation (\ref{expanded_caf}) which is not necessarily fully linear. While this may be the case, for systems of the form of equation (\ref{expanded_caf}), where the control term is state dependent, a common extension to the linear optimal control problem is to solve the state-dependent Riccati equation. This is equivalent to re-solving the Ricatti equation online at each point in time. This effectively gives us a trade-off: on one hand, this is a far easier problem to solve in comparison to the optimal control problem for a nonlinear system but not quite as simple as that for a fully linear system. This is a satisfactory trade-off and as such is the approach adopted in the current work to solve the optimal control problems for the examples considered in section \ref{results}. For the interested reader, a more detailed discussion on the state-dependent Riccati equation can be found in \cite{Kaiser2021} and \cite{Nekoo2019}. 

In this case, as discussed by the authors of \cite{Kaiser2021}, by solving the state-dependent Riccati equation the resulting control law may be interpreted as a gain-scheduled controller: 

\begin{equation}\label{gain_scheduled_controller}
\mathbf{u} = - \mathbf{K}(\mathbf{z})\mathbf{z}
\end{equation}
where in this case, the gain matrix is a function of the state, $\mathbf{K}: \mathbb{R}^{n_z} \to \mathbb{R}^{n_u} \times \mathbb{R}^{n_z}$. The control law, equation \ref{gain_scheduled_controller}, is then applied directly to the original system. Notice that the control law is linear in the new state, $\mathbf{z}$, but nonlinear in the original state, $\mathbf{x}$, given the transformation imposed by the neural network, $\mathbf{z}  = \mathbf{N(x;p)}$. 

Finally, it should be noted that when solving the optimal control problem for the new system representation, it is important to ensure that the reference being tracked in the new system corresponds to the desired reference in the original system. For the proposed framework in this paper, this can be done by passing the original reference through the network: $\mathbf{z_{ref}} = \mathbf{N(x_{ref};p)}$. This subtlety is again noted in section \ref{results}.

\section{Numerical Examples: Results and Discussion}\label{results}
In this section, we apply the proposed deep learning framework discussed in section \ref{nonlinear_systems_w_inputs} to two examples. In section \ref{lit_eg}, we consider a commonly cited example from the literature \cite{Proctor2018}, \cite{Kaiser2021}. In section \ref{CSTR_eg}, we consider the Continuous Stirred Tank Reactor system. Additionally, in section \ref{challenges}, we discuss some challenges posed by the deep learning framework, namely overfitting and the curse of dimensionality. Finally, in section \ref{future}, we summarise future research directions in the context of the challenges discussed throughout this section. 

\subsection{An Example from the Literature} \label{lit_eg}

Consider the two state control-affine system defined below:

\begin{equation}\label{simple_sys}
\begin{aligned}
  \dot{x}_1 &= \mu x_1 \\
  \dot{x}_2 &= \lambda (x_2-x_1^2) + u
\end{aligned}
\end{equation}

We seek a transformation of state $\mathbf{x}$, such that we satisfy equation (\ref{expanded_caf}). Before applying the deep learning framework to solve this problem, we note that a closed form solution exists for this system. Specifically, by defining the below transformation:

\begin{equation}\label{simple_trans}
\begin{aligned}
  z_1 &= x_1 \\
  z_2 &= x_2 \\
  z_3 &= x_1^2
\end{aligned}
\end{equation}

One can easily verify that the dynamics of the new state, $\mathbf{z}$, satisfy equation (\ref{expanded_caf}):

\begin{equation}\label{simple_trans_dynamics}
\begin{bmatrix}
\dot{z_1}\\
\dot{z_2}\\
\dot{z_3}\\
\end{bmatrix}
=
\underbrace{\begin{bmatrix}
\mu & 0 & 0\\
0 & \lambda & -\lambda \\
0 & 0 & 2\mu
\end{bmatrix}}_{\mathbf{A}}
\begin{bmatrix}
z_1\\
z_2\\
z_3
\end{bmatrix}
+
\underbrace{\begin{bmatrix}
0\\
1\\
0
\end{bmatrix}}_{\mathbf{J_z(x)B}}
u
\end{equation}

Notice that, in this specific case, due to the defined transformations in equation (\ref{simple_trans}), the control term, $\mathbf{J_z(x)B}$, in equation (\ref{simple_trans_dynamics}) is not state-dependent. Therefore, equation (\ref{simple_trans_dynamics}) defines a fully linear system and the optimal control problem amounts to solving the LQR problem. However, this will not generally be the case. 

In fact, for the proposed deep learning framework, as the transformation $\boldsymbol{\phi}$, is learned from data, and is thus purely dependent on the neural network, then it is unlikely that the control term will be state-independent. In this regard, a major challenge, which is an important area of future investigation, is overfitting. Indeed, this is an issue with all data-driven approaches to the solution of the Koopman operator \cite{Kaiser2021}. Specifically, in the case of a deep learning framework, with such a highly parametric model, it is likely that overfitting can occur with spurious information being learned from the data. In this case, although the dynamics of the original system may be faithfully captured by satisfaction of equation (\ref{chain_caf}), this can be potentially achieved with many different transformations and thus many different $\mathbf{A}$ matrices so long as they minimise the objective function within the range of the training data. Consequently, although the resulting transformation may be valid within the range of the training data, it may be a poor representation outside of this range, as will be demonstrated in section \ref{CSTR_eg} and \ref{challenges}. As a result, validation of the learned transformation is a critical step which has been discussed in various other data-driven Koopman approaches \cite{Kaiser2021}.

For the proposed deep learning framework, this is no different and overfitting can be avoided in many ways ranging from increasing the amount of data collected, cross-validation to general regularization methods \cite{Courville2016}. However, as will be demonstrated by the results throughout this section, as long as the resulting learned model is only used within the bounds of the training data, then the model is valid. In this respect, we can regard the proposed framework as learning a linearized model, in the form of equation (\ref{expanded_caf}), which has a broader range of validity than a model linearized via standard methods. Therefore, in the current work, we ensure that the data used for training encompasses a broad range such that the resulting model is valid within the bounds of the training data. These ideas are critically explored and analysed in further detail in section \ref{challenges}.

We now apply the deep learning framework to the system defined in equation (\ref{simple_sys}). The objective is to find a transformation of the original system which results in a system of the form of equation (\ref{expanded_caf}). By identifying such a system, we will solve the optimal control problem, as discussed in section (\ref{solve_ocp}), to regulate the states to the origin. Note that for the simulations performed in this paper, state one, $x_1$, was chosen to be a stable state with $\mu  = -0.1$, while state two, $x_2$, was defined to be unstable with $\lambda = 1$. Similar to the exact solution, defined in equation (\ref{simple_trans}), we fix the new system dimension to three. Additionally, as the intention is to control states $x_1$ and $x_2$, then we specify that the first two states of the new system are identical i.e. $z_1 = x_1$ and $z_2 = x_2$. We allow the final state, $z_3$, to be freely chosen by the neural network. However, this is not necessary and instead we can make all states free to be chosen by the neural network. In such a case, as discussed in section (\ref{solve_ocp}), it is important to ensure that the reference being tracked in the new system corresponds to the desired reference in the original system. This can be done by passing the original reference through the network: $\mathbf{z_{ref}} = \mathbf{N(x_{ref};p)}$.

\begin{figure}
\begin{center}
\includegraphics[width=0.8\textwidth]{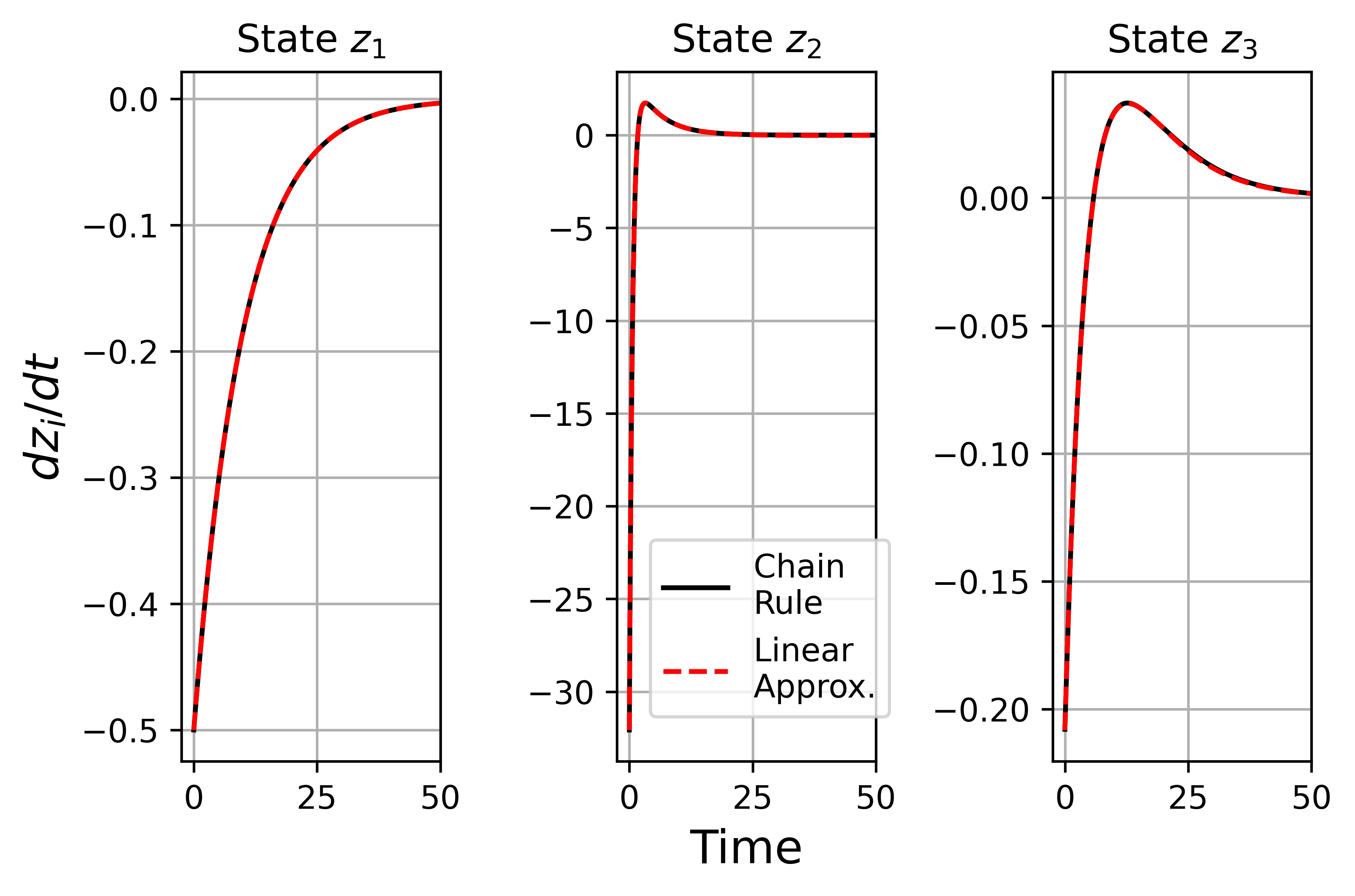} 
\caption{
Results of the optimization showing a comparison between the exact time derivative calculated via the chain rule (black line) i.e. $\frac{\mathrm{d}\mathbf{z}}{\mathrm{d}t} = \mathbf{J_z(x)}\frac{\mathrm{d}\mathbf{x}}{\mathrm{d}t}$ and the linear approximation (red dashed line) i.e. $\frac{\mathrm{d}\mathbf{z}}{\mathrm{d}t} = \mathbf{A}\mathbf{z} + \mathbf{J_z(x)}\mathbf{Bu}$. 
} 
\label{time_deriv_compar}
\end{center}
\end{figure}

Finally, data was collected from equation (\ref{simple_sys}) by simulation of a number of different trajectories. This data was then used to solve the optimization problem, equation (\ref{optimisation_control_sum}). Fig. \ref{time_deriv_compar} shows the results of the optimization for a given training trajectory. The black curves represent the exact time derivative calculated from the chain rule i.e. $\frac{\mathrm{d}\mathbf{z}}{\mathrm{d}t} = \mathbf{J_z(x)}\frac{\mathrm{d}\mathbf{x}}{\mathrm{d}t}$, while the red dashed curves are the results of the linear approximation i.e. $\frac{\mathrm{d}\mathbf{z}}{\mathrm{d}t} = \mathbf{A}\mathbf{z} + \mathbf{J_z(x)}\mathbf{Bu}$. It will be noticed that these are in good agreement indicating that a successful model has been identified within the bounds of the training data. The results for the rest of the data were similar and this is discussed in further detail in section \ref{challenges}.

\begin{figure}
\begin{center}
\includegraphics[width=0.8\textwidth]{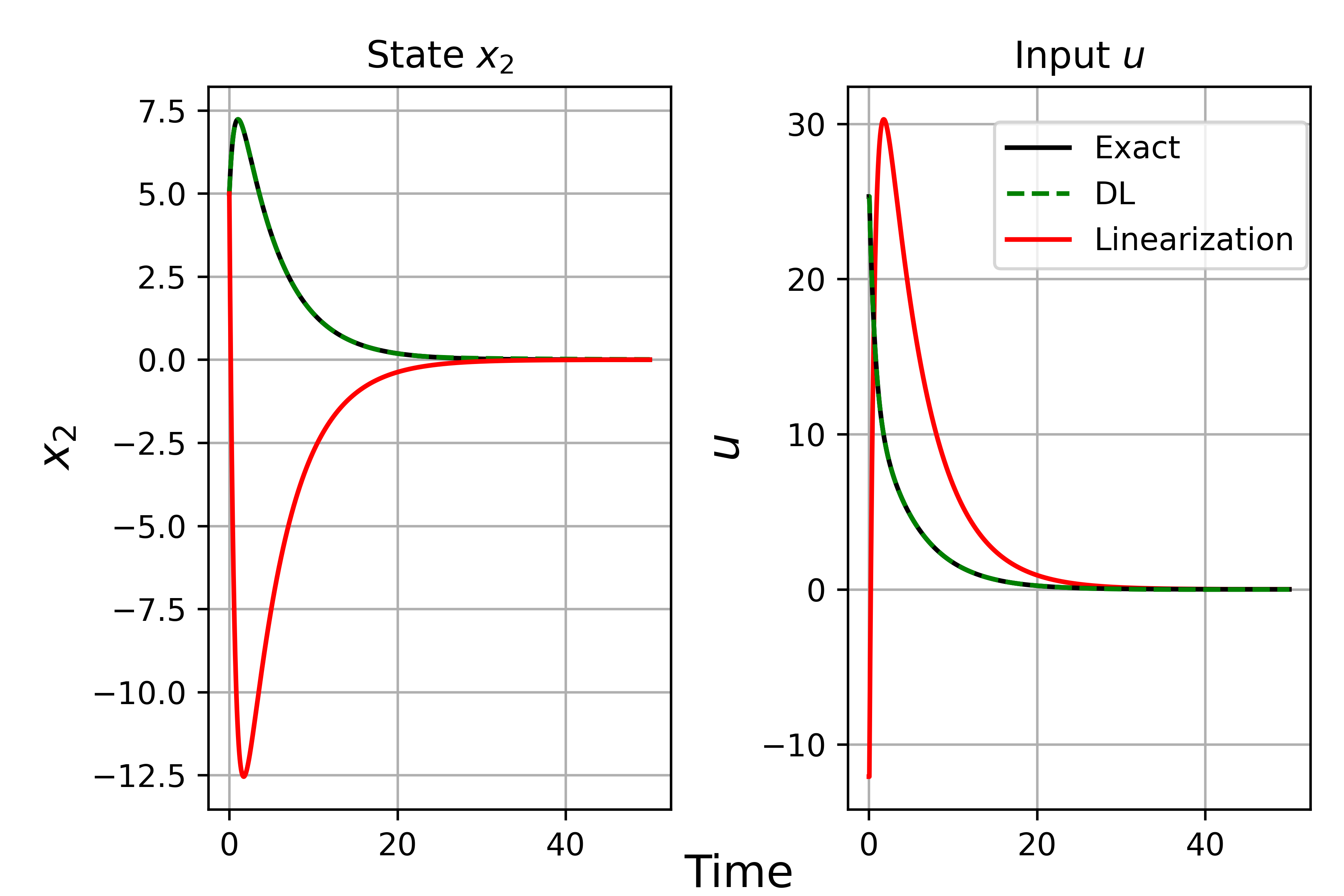} 
\caption{
Results of optimal control applied to the full nonlinear system with the controller designed using a model obtained by standard linearization (red line), the exact solution (black line) from equation (\ref{simple_trans_dynamics}) and the model obtained from the deep learning framework (green dashed line).
} 
\label{lqr_control}
\end{center}
\end{figure}

The resulting system, of the form of equation (\ref{expanded_caf}), was then used to solve the optimal control problem to control the states to the origin from an initial condition of $(x_1,x_2)=(5,5)$. The results of this can be seen in Fig. \ref{lqr_control} for the unstable state, $x_2$, and the control action, $u$. In both figures, the red line represents the results of the controller designed using the model linearized around the origin, the black line represents the results when using the exact solution, equation (\ref{simple_trans_dynamics}), while the green dashed line represents the results when using the model from the proposed deep learning framework. It is clear to see that the proposed framework (green dashed line) performs equivalently to the exact solution (black line) by tightly controlling the states to the origin. In comparison, the controller designed using the model obtained by standard linearization (red line) shows poor, sluggish performance as the model is only valid close to the origin.  However, it is important to note that while these results are promising, this is only within the bounds of the training data.  To further exemplify this point, notice that the learned $\mathbf{A}$ matrix, which is valid only within the bounds of the training data, shown in equation (\ref{learned_A}) is noticeably different from the exact $\mathbf{A}$ matrix in equation (\ref{simple_trans_dynamics}), which is valid for the whole domain. Despite this observation, the proposed deep learning framework has effectively learned a linearized model which has a larger range of validity in comparison to the model obtained by standard linearization. This is further discussed in section \ref{challenges}.

\begin{equation}\label{learned_A}
\mathbf{A_{learned}}
=
\begin{bmatrix}
-0.10 & 0.00 & 0.00\\
-5.50 & 1.01 & -12.77\\
-0.05 & 0.00 & -0.21\\
\end{bmatrix}
\end{equation}

\subsection{The Continuous Stirred Tank Reactor (CSTR)}\label{CSTR_eg}
In this section, we apply the proposed deep learning framework to a CSTR example. We restrict ourselves to the single state control-affine system defined below:

\begin{equation}\label{cstr_sys}
\frac{\mathrm{d}T}{\mathrm{d}t} = \frac{q}{V}(T_i - T) -\frac{\Delta H_R}{\rho C_p}ke^{\frac{-E_A}{RT}}C_A + \frac{UA}{\rho C_p V}(T_c - T)
\end{equation}

where $T$ is the temperature of the reactor and the only state of the system while $T_c$ represents the cooling water temperature which is the input to the system. All other variables are treated as fixed parameters. Data was collected from the system by simulation of equation (\ref{cstr_sys}). This was done by first finding a fixed point for the system where $\frac{\mathrm{d}T}{\mathrm{d}t} = 0$, so as to identify steady-state values of the reactor temperature and cooling water temperature which we denote by $T_{ss}$ and $T_{c,ss}$ respectively. The system was then simulated by defining a series of random step inputs for the cooling water temperature between lower and upper bounds to generate a diverse dataset within the training data range. Consequently, the resulting reactor temperature data and its time derivative were collected throughout the length of the simulation. This data was then used to solve the optimization problem, equation (\ref{optimisation_control_sum}). The number of new system dimensions were ranged between two to five, however, beyond this, training was slower due to the need for more data. 

\begin{figure}[t]
\begin{center}
\includegraphics[width=0.8\textwidth]{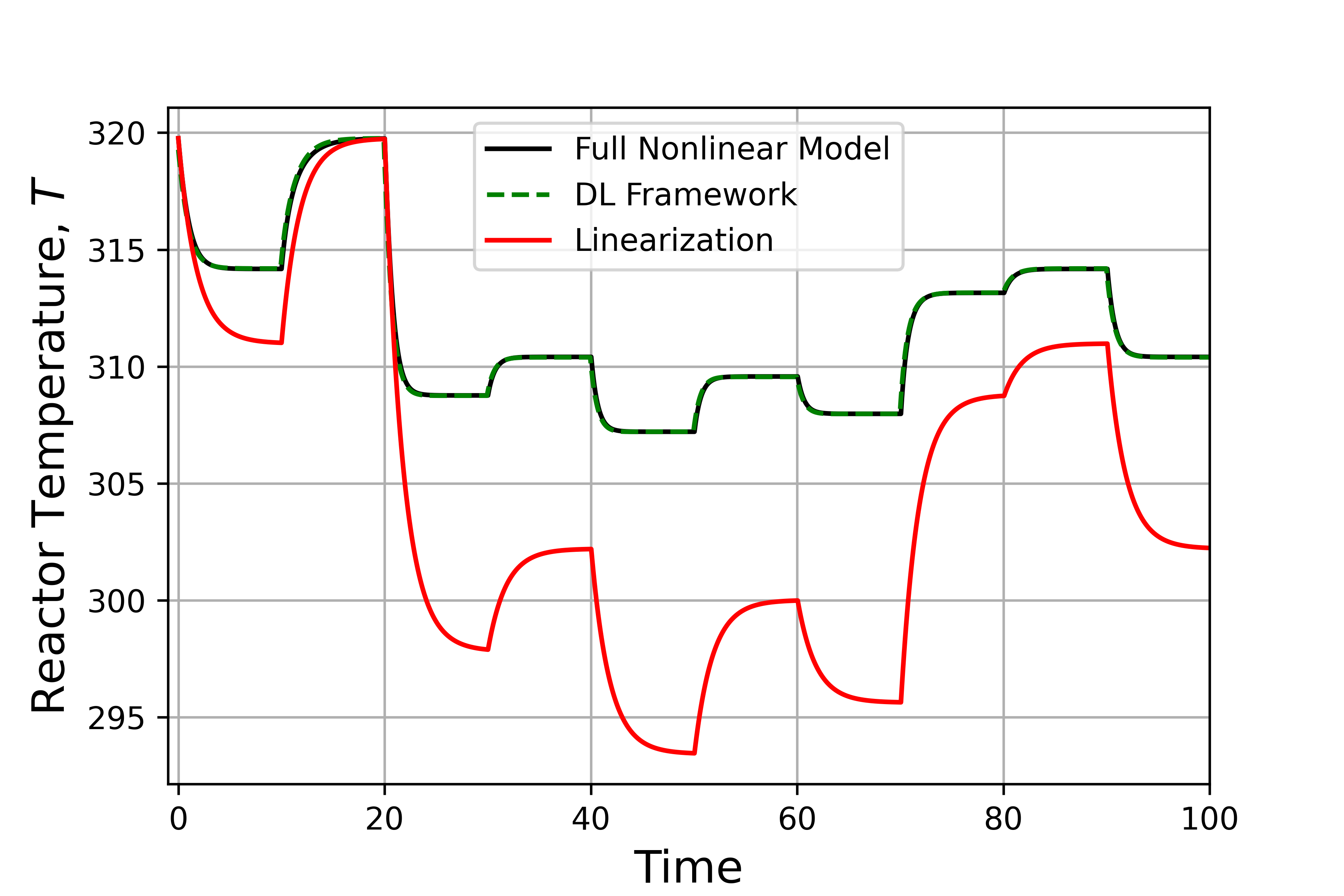} 
\caption{
Simulation results of the full nonlinear model (black line), the model obtained from the deep learning framework (green dashed line) and the model obtained from standard linearization around the fixed point (red line) when the system was perturbed with random step inputs within the training range.
} 
\label{sim_in_range}
\end{center}
\end{figure}

To assess the quality of the learned models, the new system representations, of the form of equation (\ref{expanded_caf}), were simulated for various new random step inputs to the system, while staying within the bounds of the training data. Fig. \ref{sim_in_range} shows the results of these simulations for a single example where the new system dimension was three. Note that the results were similar across the different cases of lifted dimensions. The black line represents the simulation results of the full nonlinear model, equation (\ref{cstr_sys}), while the green dashed line represents the results from the model learned by the deep learning framework. As can be seen, the learned linear representation accurately captures the original system dynamics. For comparison, using standard methods, the CSTR model was linearized around the fixed point, $(T_{ss},T_{c,ss})$ and was also simulated for the same step inputs. The results of this simulation are represented by the red line. It is clear to see that the linearized model performs poorly far away from the point of linearization, $(T_{ss},T_{c,ss})$, due to the mismatch with the full nonlinear system.

\begin{figure}
\begin{center}
\includegraphics[width=0.8\textwidth]{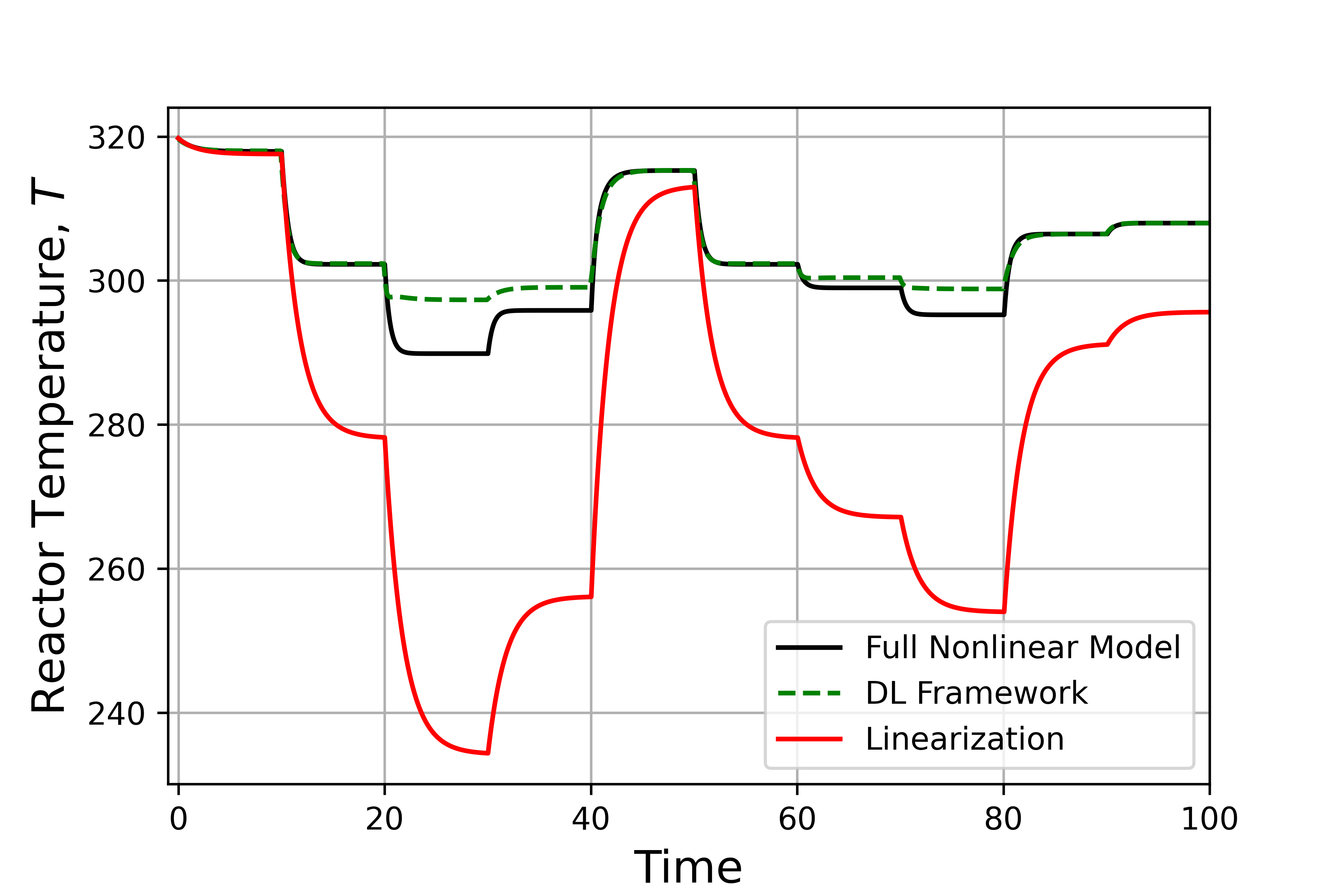} 
\caption{
Simulation results of the full nonlinear model (black line), the model obtained from the deep learning framework (green dashed line) and the model obtained from standard linearization around the fixed point (red line) when the system was perturbed with random step inputs which were allowed to exceed the training range.
} 
\label{sim_out_range}
\end{center}
\end{figure}

In addition to this, Fig. \ref{sim_out_range} shows the results of the same system when new random step inputs to the system were allowed to exceed the bounds of the training data. In this case, it can be seen that the learned linear representation generated by the deep learning framework (green dashed line) does a poor job of capturing the original system dynamics (black line) outside of the training range. 

\begin{figure}
\begin{center}
\includegraphics[width=0.8\textwidth]{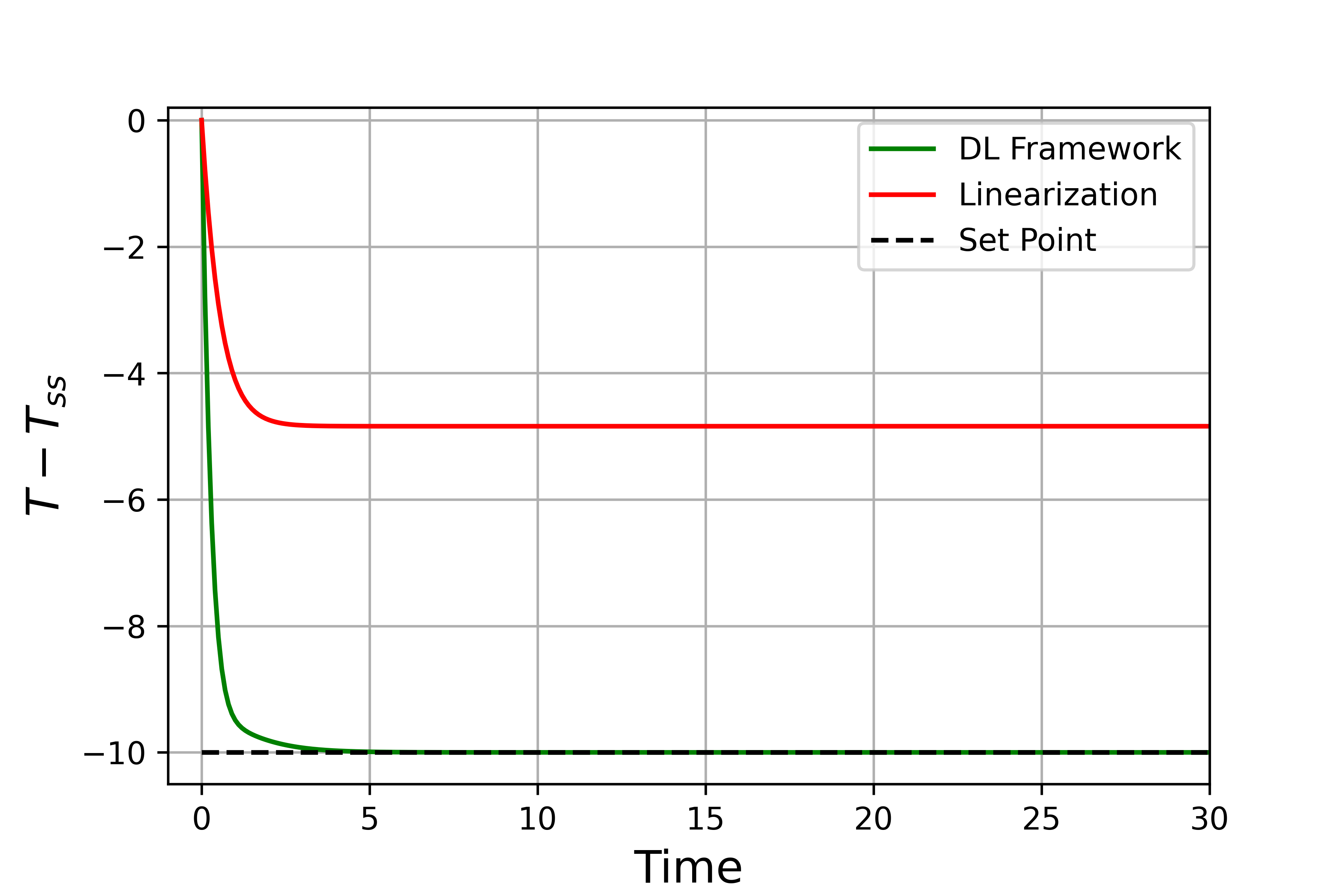} 
\caption{
The model obtained by standard linearization (red line) and the model obtained from the deep learning framework (green line) were used to design a controller to track the set-point: $T_{sp} = T_{ss} - 10$. The resulting control actions were then applied to the full nonlinear system giving the results portrayed in their respective colours.
} 
\label{cstr_control}
\end{center}
\end{figure}

In the last set of numerical experiments, the learned linear representations and the standard linearized model were used to design a controller to track the set-point: $T_{sp} = T_{ss} - 10$, with initial conditions starting at the fixed point, $(T_{ss},T_{c,ss})$. The set-point was chosen to be within the range of the training data, however, it is far away from the point where the linearized model is valid as demonstrated by Fig. \ref{sim_in_range} and Fig. \ref{sim_out_range}. As a result of this, as can be seen in Fig. \ref{cstr_control}, a controller was designed to successfully control the reactor temperature to the new set-point when using the learned linear representation of the full nonlinear system (green line). Alternatively, a controller designed using the linearized model fails to track the set-point due to the mismatch with the full nonlinear system (red line). 

\subsection{Overfitting and The Curse of Dimensionality}\label{challenges}

As discussed throughout section \ref{lit_eg} and \ref{CSTR_eg}, the issue of overfitting is a major challenge for all data-driven solutions to the Koopman operator, with the proposed deep learning framework not being any different. In this section, we will analyse and discuss this concept in more detail by demonstrating the challenges posed by the issue of overfitting and how this can be used to our advantage. 

We start this analysis by restricting our attention to the motivating example from section \ref{motivating_example}. We do this as we know a closed from solution exists for the transformation (equation \ref{simple_transformation}) which converts the original system (equation \ref{simple_example}) into a linear system (equation \ref{simple_linear}). Additionally, as the defined transformation from the original state, $x$, to the new state, $z$, stays within the same single dimension i.e. $x,z \in \mathbb{R}$, we can use this fact to easily visualize the results which helps to solidify the discussed concepts. 

While a closed form solution exists for this system, as defined by equation (\ref{simple_transformation}), we can use the proposed deep learning framework to investigate the transformation which is learned by the neural network. In this case, as the system has no inputs, the optimization problem defined in equation (\ref{optimisation_sum}) is solved. We deliberately restrict the training data range to the closed interval: $x^{(i)} \in [0,5]$. This is so that we can compare the fidelity of the learned transformation inside and outside of the training data range. 

\begin{figure}
\begin{center}
\includegraphics[width=0.8\textwidth]{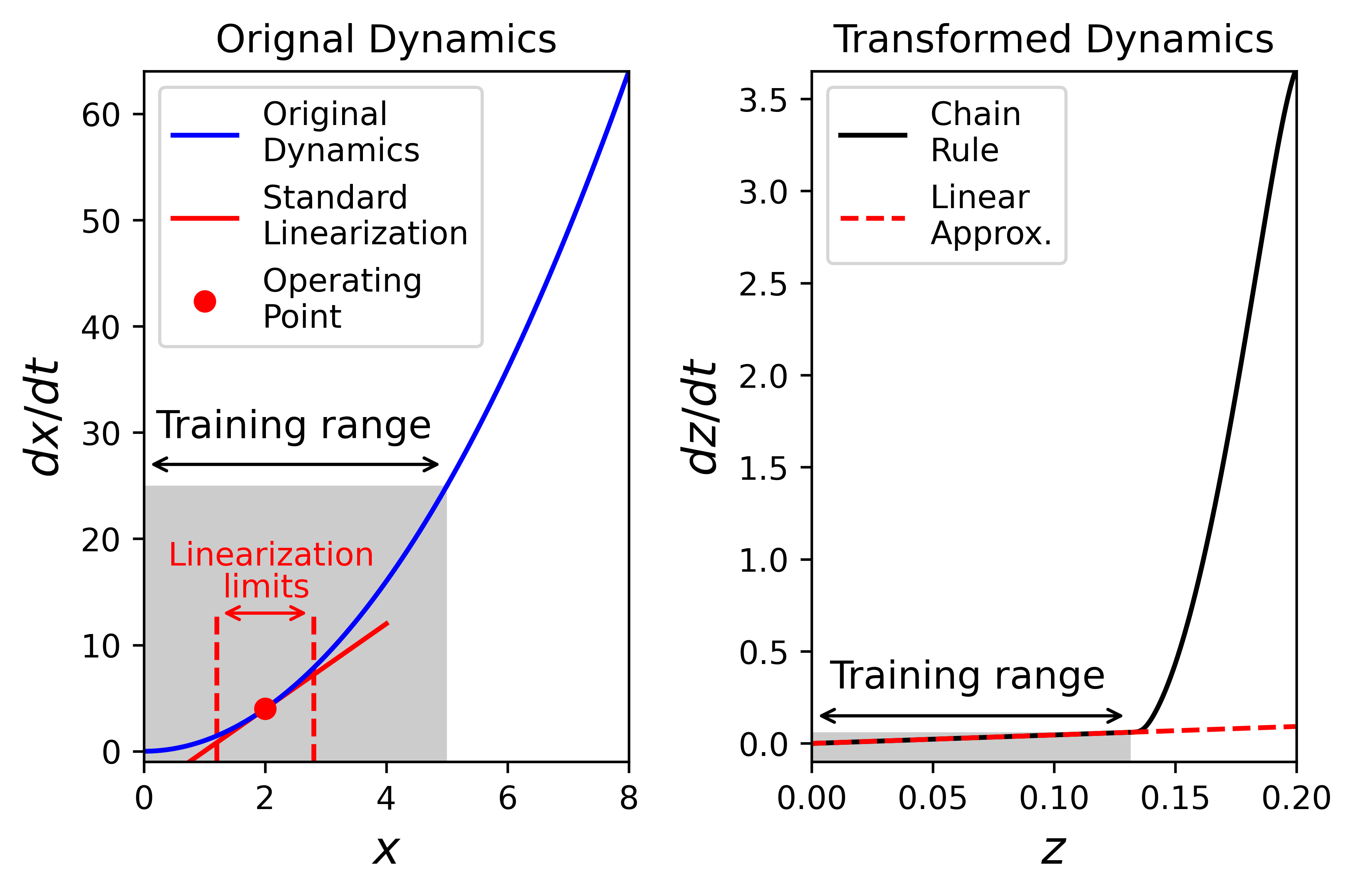} 
\caption{
The results of the deep learning framework applied to the motivating example from section \ref{motivating_example} defined by equation (\ref{simple_example}). On the left hand subplot, the blue curve defines the original system dynamics. The red line defines a linear approximation, obtained using standard methods, at the red operating point while the red dashed vertical lines define the lower and upper bounds within which the linear approximation is valid. The grey region defines the training data range for the deep learning framework. On the right hand subplot, the results of the optimization are shown giving a comparison between the exact time derivative calculated via the chain rule (black line) and the linear approximation (red line) inside and outside of the training range in grey. 
} 
\label{motivating_eg}
\end{center}
\end{figure}

The results of the optimization are given in Fig. \ref{motivating_eg}. On the left hand subplot, the blue curve represents the original system dynamics, as defined by equation (\ref{simple_example}). The red line defines a linear approximation, obtained using standard methods, at the red operating point. The red dashed vertical lines define the lower and upper bounds for which the standard linear approximation is valid within. The grey region defines the training data range used for the proposed framework.  On the right hand subplot are the result of the optimization i.e. the new system dynamics based on the learned transformation. As before, the grey region defines the results within the training data range while the white region defines the results outside of this range. The black line represents the exact time derivative calculated from the chain rule i.e. $\frac{\mathrm{d}\mathbf{z}}{\mathrm{d}t} = \mathbf{J_z(x)}\frac{\mathrm{d}\mathbf{x}}{\mathrm{d}t}$, while the red dashed line is the result of the learned linear approximation i.e. $\frac{\mathrm{d}\mathbf{z}}{\mathrm{d}t} = \mathbf{A}\mathbf{z}$. It is clear from the results that these are both in good agreement within the training data range suggesting that a successful model has been identified within this range. However, outside of the training range, the two curves diverge i.e. the learned linear approximation given by $\frac{\mathrm{d}\mathbf{z}}{\mathrm{d}t} = \mathbf{A}\mathbf{z}$ does not satisfy the original system dynamics as defined by the chain rule, $\frac{\mathrm{d}\mathbf{z}}{\mathrm{d}t} = \mathbf{J_z(x)}\frac{\mathrm{d}\mathbf{x}}{\mathrm{d}t}$. Therefore, the learned linear approximation is no longer valid outside of the training data range. 

These results highlight the key points which were made regarding overfitting discussed throughout section \ref{lit_eg} and \ref{CSTR_eg}. Specifically, the proposed framework learns a linearized model which can be used validly as long as the system remains within the training data range.  However, outside of this range, without any strategies for dealing with overfitting, no guarantees can be made. Despite this, as can be seen in Fig. \ref{motivating_eg}, the learned linear model from the proposed framework has a larger range of validity (the training range in grey) than a model obtained via standard linearization (the red dashed vertical bounds on the left hand subplot).

As a final note, for this specific system, if we wanted to increase the range of validity for the learned linear model then this can be easily done by simply increasing the training data range. However, it should be noted that, while extending the training data range for this single dimensional system may be trivial, this no longer becomes an easy task for higher dimensional systems where the curse of dimensionality becomes an issue to contend with \cite{Bellman2003}, \cite{Verleysen2005}. This challenge is discussed in further detail later in this section. 

\begin{figure}
\begin{center}
\includegraphics[width=0.8\textwidth]{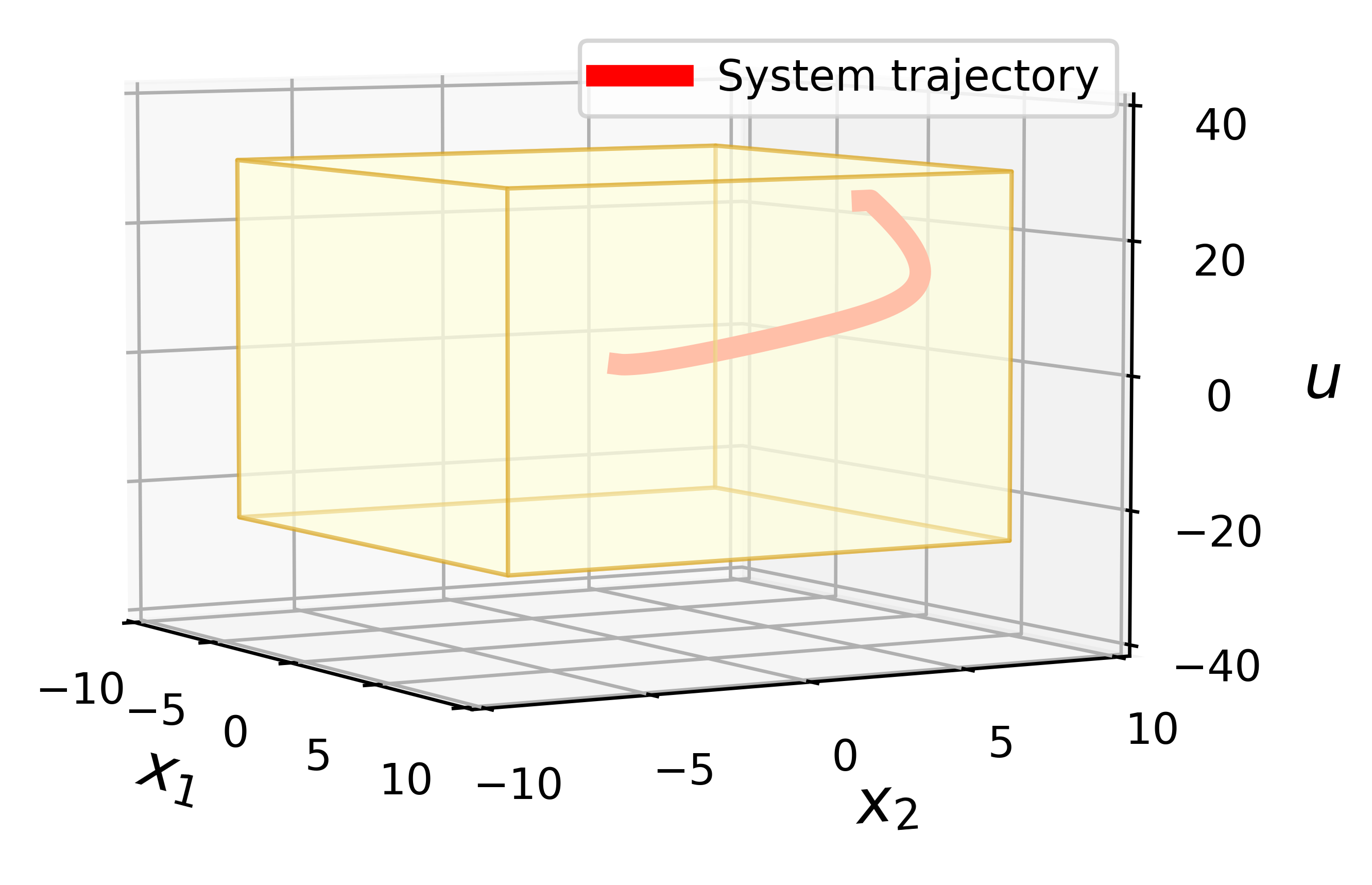} 
\caption{
A subset of state-space for the system defined by equation (\ref{simple_sys}). The yellow cuboid defines the bounds of the training data and the red curve defines the system trajectory as per the results of Fig. \ref{lqr_control}. In this case, the system stays within the region in which the learned linear model is valid i.e. the yellow cuboid. 
} 
\label{in_bounds}
\end{center}
\end{figure}

The above arguments can also be extended to higher dimensional systems. To demonstrate this, we now reconsider the system from section \ref{lit_eg} defined by equation (\ref{simple_sys}). Fig. \ref{in_bounds} shows a subset of state-space for the defined system. The yellow cuboid defines the bounds of the training data from section \ref{lit_eg} used to solve the optimization problem defined in equation (\ref{optimisation_control_sum}). The red curve defines the system trajectory when the controller designed using the learned linear model was applied to the system, as per the green dashed curves shown in Fig. \ref{lqr_control}. Notice that the system trajectory stays within the yellow cuboid i.e. the region in which the learned linear model is valid. As the system remains within this region, the model can be used with confidence. This results in the successful control of the system as depicted by the green dashed curves in Fig. \ref{lqr_control}.

\begin{figure}
\begin{center}
\includegraphics[width=0.8\textwidth]{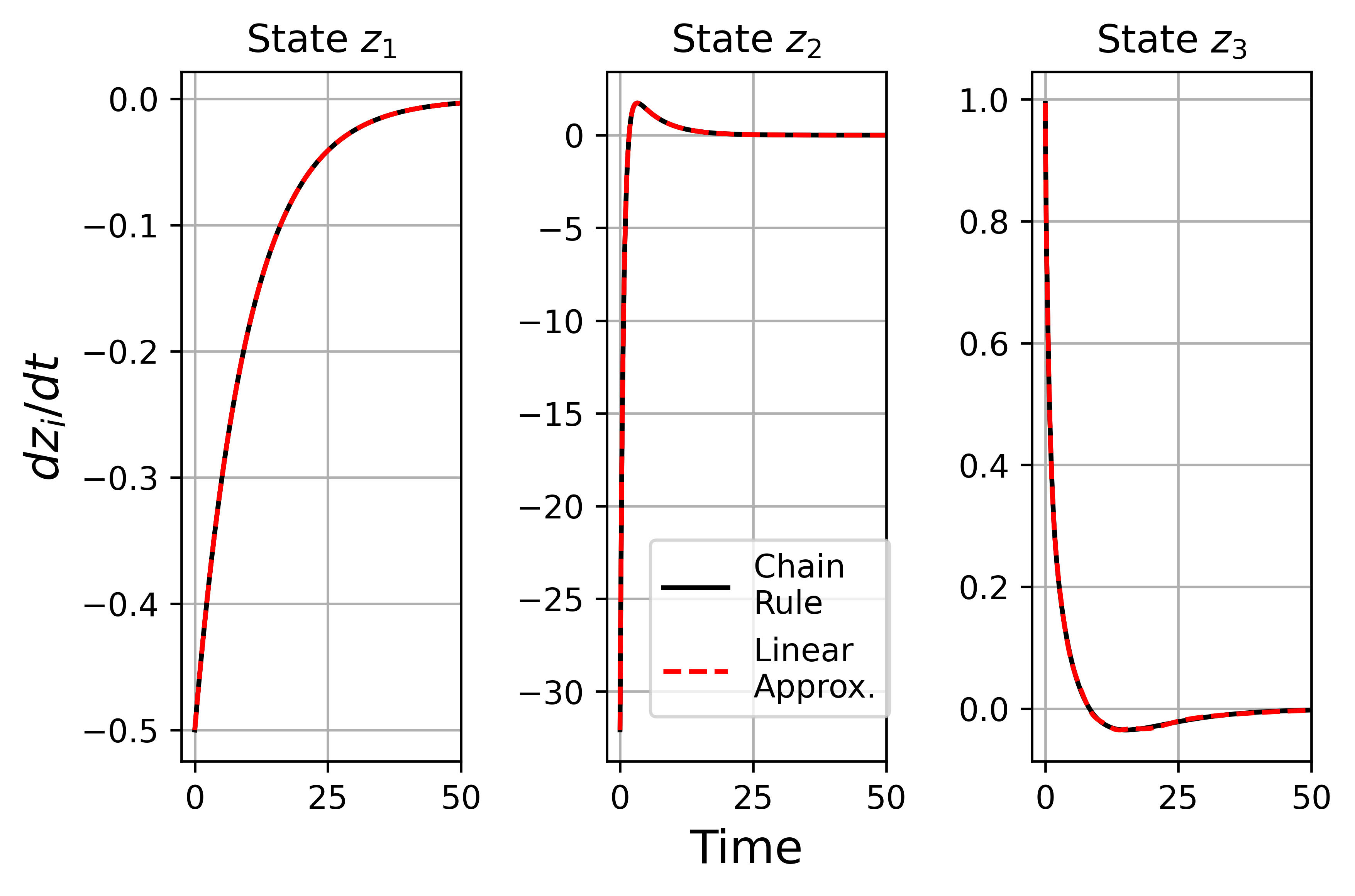} 
\caption{
Results of the optimization showing a comparison between the exact time derivative calculated via the chain rule (black line) i.e. $\frac{\mathrm{d}\mathbf{z}}{\mathrm{d}t} = \mathbf{J_z(x)}\frac{\mathrm{d}\mathbf{x}}{\mathrm{d}t}$ and the linear approximation (red dashed line) i.e. $\frac{\mathrm{d}\mathbf{z}}{\mathrm{d}t} = \mathbf{A}\mathbf{z} + \mathbf{J_z(x)}\mathbf{Bu}$.
} 
\label{time_deriv_compar_poor}
\end{center}
\end{figure}

\begin{figure}
\begin{center}
\includegraphics[width=0.8\textwidth]{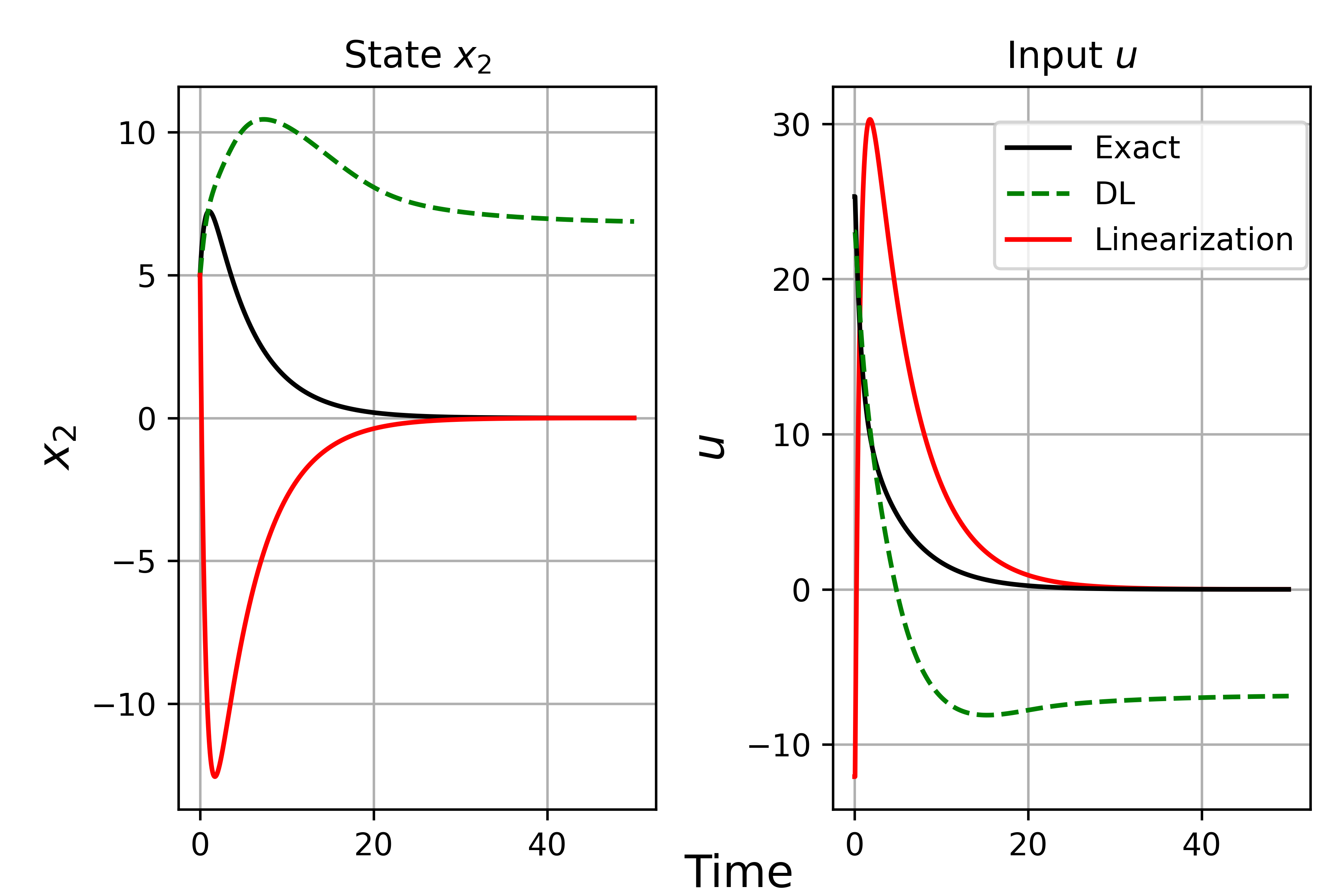} 
\caption{
Results of optimal control applied to the full nonlinear system with the controller designed using a model obtained by standard linearization (red line), the exact solution (black line) from equation (\ref{simple_trans_dynamics}) and the model obtained from the deep learning framework (green dashed line). Notice that, in this case, the model learned by the deep learning framework was trained with a limited dataset resulting in the poor control of the system as depicted by the green dashed line. 
} 
\label{poor_control}
\end{center}
\end{figure}

Alternatively, we can investigate the consequences of using a limited and narrower dataset to learn the linearized model. In the first instance, Fig. \ref{time_deriv_compar_poor} shows the results of the optimization for a given training trajectory. Similar to the results in Fig. \ref{time_deriv_compar}, it appears that a model has been successfully identified within the bounds of the training data. However, in this case, when the optimal control problem is solved using the learned linear model and the resulting controller is applied to the original system, this results in the poor control of the system. This is depicted in Fig. \ref{poor_control}. Similar to Fig. \ref{lqr_control}, the black line represents the results when using the exact solution, equation (\ref{simple_trans_dynamics}), while the green dashed line represents the results when using the model from the proposed deep learning framework. Similarly, the red line shows the results when using the model obtained by standard linearization. In this case, it is clear that the controller designed using the learned linear model (green dashed line) performs poorly even in comparison to the model obtained from standard linearization (red line). In fact, in this case, the controller fails to regulate the unstable state, $x_2$, to the origin.

\begin{figure}
\begin{center}
\includegraphics[width=0.8\textwidth]{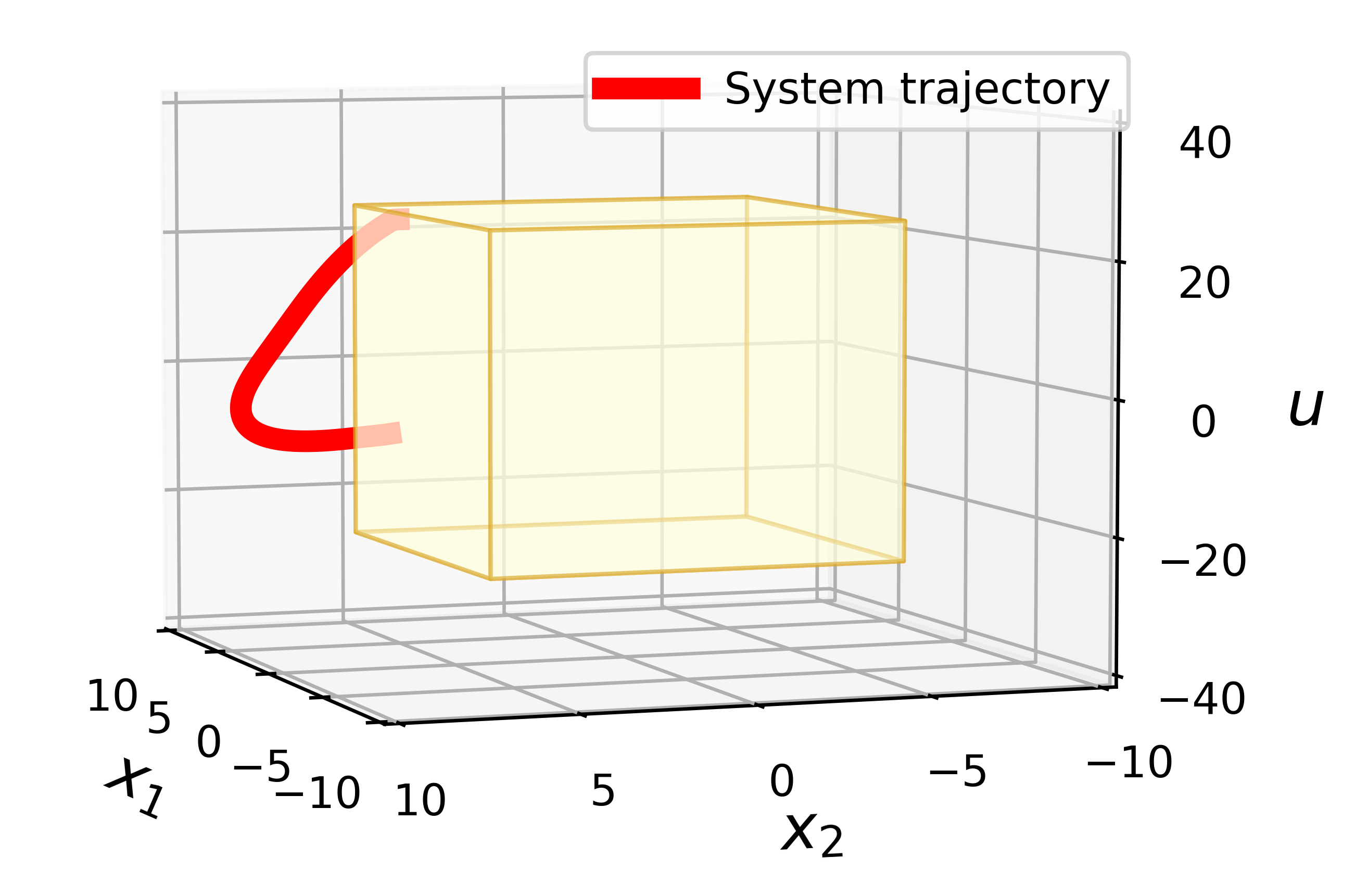} 
\caption{
A subset of state-space for the system defined by equation (\ref{simple_sys}). The yellow cuboid defines the bounds of the training data and the red curve defines the system trajectory as per the results of Fig. \ref{poor_control}. In this case, the dataset used to learn the linearized model is narrower and more limited than that from Fig. \ref{in_bounds}, and consequently the system trajectory moves outwith the region in which the learned linear model is valid i.e. the yellow cuboid. 
} 
\label{out_bounds}
\end{center}
\end{figure}

The reason for this becomes immediately clear when we consider the system trajectory with respect to the bounds of the new training data. This is depicted in Fig. \ref{out_bounds}. As before, the yellow cuboid defines the bounds of the training data i.e. the region in which the learned model is valid. This is a more limited and narrower dataset in comparison to the dataset from Fig. \ref{in_bounds}. The red curve defines the system trajectory when a controller, designed using the learned linear model, was applied
to the system as per the green dashed curves shown in Fig. \ref{poor_control}. In this case, it can be seen that the system moves outside of the region in which the learned linear model is valid i.e. the yellow cuboid. Although the system initially starts within the yellow cuboid, as a result of the control action, the system is steered outside of this region. As a result, outside of this region, an invalid model is being used to design the controller. This results in the poor control of the system as depicted by the green dashed curves in Fig. \ref{poor_control}. 

\begin{figure}
\begin{center}
\includegraphics[width=0.8\textwidth]{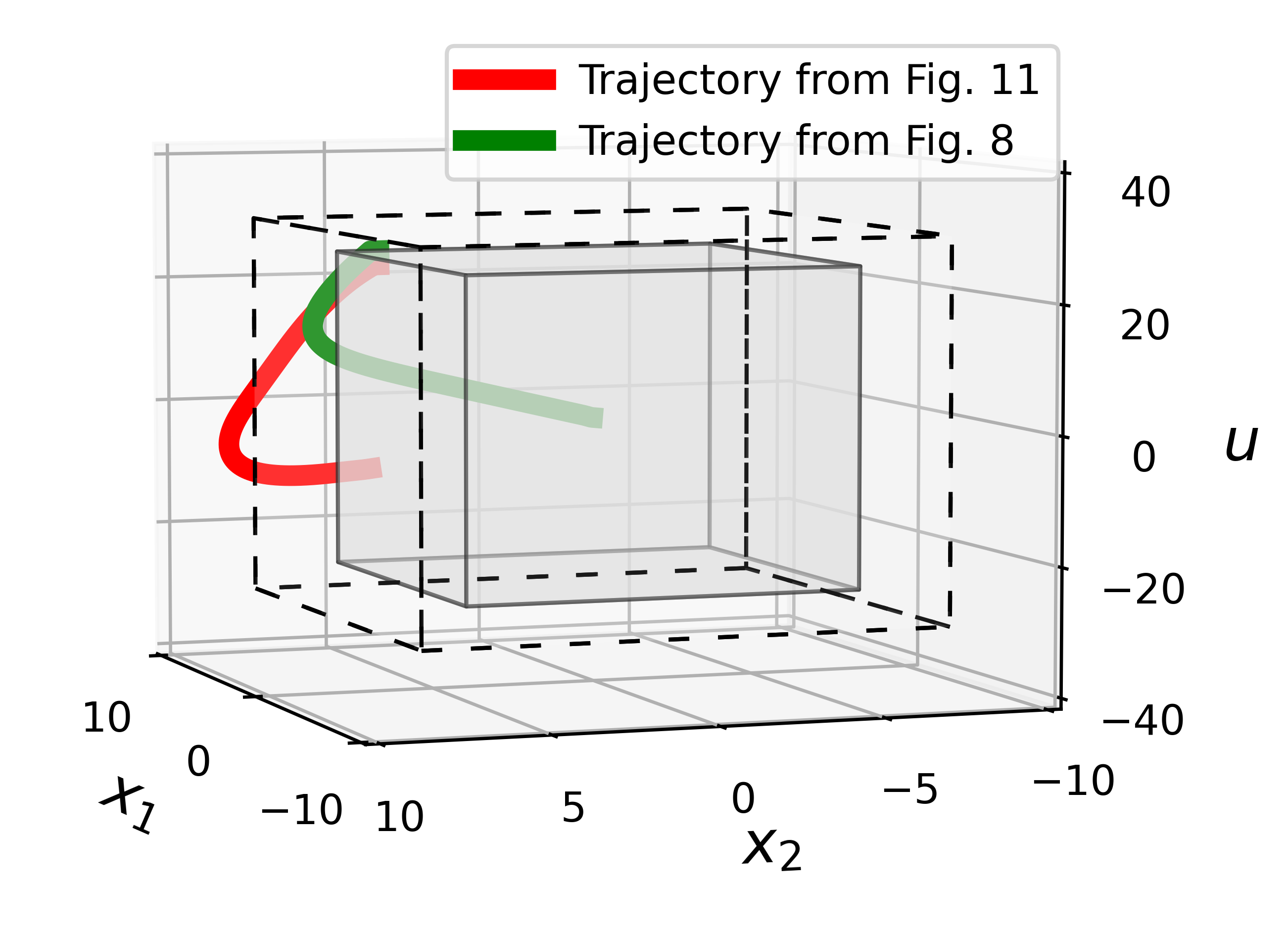} 
\caption{
A comparison of the trajectories and datasets from Fig. \ref{in_bounds} and \ref{out_bounds}. The green curve and the transparent cuboid with black dashed edges defines the system trajectory and dataset from Fig. \ref{in_bounds}. The red curve and the grey cuboid defines the system trajectory and dataset from Fig. \ref{out_bounds}.
} 
\label{sim_all}
\end{center}
\end{figure}

For comparison, Fig. \ref{sim_all} shows the system trajectories and datasets from both Fig. \ref{in_bounds} and \ref{out_bounds} in a single plot. The green curve defines the system trajectory from Fig. \ref{in_bounds} and the corresponding dataset used for training is represented by the transparent cuboid with black dashed edges. On the other hand, the red curve defines the system trajectory from Fig. \ref{out_bounds} and the corresponding dataset is represented by the grey cuboid. In this diagram, it is clear to see that the dataset from Fig. \ref{in_bounds} (transparent cuboid) is broader and larger than the dataset from Fig. \ref{out_bounds} (grey cuboid). In particular, it can be seen that, although both trajectories start at similar initial conditions, as the red curve moves outside of the region in which the corresponding learned linear model is valid (grey cuboid), the controller never manages to regulate the states to the origin due to the plant-model mismatch. In contrast, the green curve shows the states are regulated to the origin under the action of the controller as the corresponding learned linear model is an accurate representation of the system within the corresponding training range (transparent cuboid) which the system never leaves. 

The salient point to keep in mind from the aforementioned results is that a large and broad dataset is needed to effectively learn a linearized model using the proposed framework. This is so that the model can be used with confidence within the range of interest. 

However, as alluded to earlier, while this statement may appear to be benign for lower dimensional problems, such as those considered in this section, this is not the case for higher dimensional systems. The reason for this is due to the curse of dimensionality. Specifically, suppose that we define the dimensionality of a dynamical system by the sum of the number of states and the number of inputs of the system, which we denote by $n_x$ and $n_u$ respectively. Then if we were to naively sample the system uniformly with $d$ measurements along each dimension considering all possible combinations, this would require $d^{(n_x+n_u)}$ total samples. Consequently, as the number of dimensions of the system increase, the amount of data needed would scale exponentially. This is a significant challenge as most systems of practical interest are typically high dimensional, especially, the systems which are commonly studied within process systems engineering. Therefore, this is an area of active research and a challenge which we hope to address in future work. 

With the above being said, there are already many methods which could be used to alleviate the issues caused by the curse of dimensionality. Firstly, as discussed by the authors of \cite{Kuo2005}, most high dimensional systems are only artificially high dimensional in the sense that there usually exists a lower dimensional embedding of the data in the original higher dimensional space. This fact can then be used to more effectively sample from the system as opposed to the naive approach discussed in the previous paragraph where all possible combinations of the variables are considered. These ideas are also leveraged in areas such as manifold learning and dimensionality reduction \cite{Izenman2012} \cite{Maaten2009}, \cite{Sorzano2014}. This is especially relevant for physical systems where physical relationships can be used to restrict sampling of the underlying system to feasible combinations of the physical variables. As a simple example, for systems involving pressure and temperature, we might expect high pressure and high temperature  or low pressure and low temperature combinations due to the physical correlation between these variables. However, it would be rare to find inverse correlations. Knowledge such as this can be exploited to direct the sampling strategy of the underlying system. An example of such an approach can be found in \cite{Fu2018}, where the authors use basic chemical fundamentals to efficiently sample data from reaction networks which would otherwise suffer from the curse of dimensionality. It is concepts such as these which we hope to use in future work to consider scaling of the proposed framework to larger scale systems.

\subsection{Future Work}\label{future}

Finally, although we have successfully demonstrated how the proposed deep learning framework can be used to learn a higher dimensional linear representation of a nonlinear system there are still a number of future research directions. 

Firstly, as already extensively discussed, overfitting is a challenge which faces all data-driven Koopman approaches. Consequently, in the current work, the learned model is only valid within the bounds of the training data. However, as discussed, even this is promising as it suggests that a linear model can be learned by the deep learning framework which has a broader range of validity than a model obtained by standard linearization. The resulting model can then be used for a wider range of operating conditions than a standard linearized model as shown by the results throughout this section. 

In addition to this, in section \ref{challenges}, we alluded to the challenges posed by higher dimensional problems when applying the proposed framework. Specifically, the data requirements to ensure a broad enough dataset is generated for the proposed framework starts to suffer from the curse of dimensionality. However, in section \ref{challenges}, we presented possible ways these issues could be alleviated. Consequently, this is an area of research which we intend to investigate as a path for future work. 

Lastly, in the current work,  the efficient optimization of equation (\ref{optimisation_control_sum}), relied on the use of second-order optimization algorithms. This is not common for most deep learning approaches, which instead rely on first-order optimization algorithms such as gradient descent. This is because, the construction and subsequent inversion of the Hessian matrix becomes computationally intractable for large neural network architectures due to the sheer number of parameters \cite{LeCun2012}. Indeed, this was a constraining factor for architecture choice in the current study as larger networks became more computationally intensive to train. However, preliminary investigations in to the use of second-order optimizers which use a limited-memory Hessian approximation appeared to give promising results; an observation supported by the literature \cite{Bottou2018}. This is an area which we hope to explore in more detail which should also allow for a more thorough investigation into the effect of architecture choice on the proposed deep learning framework. 

\section{Conclusions}\label{conclusions}

In this paper, we have demonstrated how deep learning can be used to learn a linearized model of a nonlinear system which has a broader range of validity than a model obtained by standard linearization. We give strong evidence that the learned model accurately captures the dynamics of the full nonlinear system within the range of the training data. Additionally, we demonstrated that the resulting learned model can then be used to successfully control the full nonlinear system. Consequently, the proposed framework could be used to model complex, nonlinear dynamical systems with higher dimensional linear representations, thus allowing for the tractable control of such systems. This is an important endeavour given that most systems of interest are distinguished by nonlinear behaviour. While these results are encouraging, many challenges remain ranging from a systematic investigation into overfitting to efficient training of the underlying neural network.  
\bibliography{koopman}

\end{document}